\documentclass[10pt]{article}

\usepackage{fontspec}
\usepackage{unicode-math}

\AtBeginDocument{\renewcommand{\mathrm}[1]{{\symup{#1}}}}

\usepackage[a4paper, margin=2.5cm]{geometry}
\usepackage{setspace}
\usepackage{xcolor}
\usepackage{ulem}
\usepackage{comment}

\newcommand{\delete}[1]{}

\newcommand{\mnote}[1]{}

\usepackage{amsmath}
\usepackage{bm}
\usepackage{siunitx}

\usepackage{graphicx}
\usepackage{booktabs}
\usepackage{multirow}
\usepackage{makecell}
\usepackage{colortbl}
\usepackage{adjustbox}
\usepackage{rotating}
\usepackage{tabularx}

\usepackage{hyperref}
\usepackage{authblk}
\usepackage{lineno}
\usepackage{textcomp, gensymb}
\usepackage[title]{appendix}
\usepackage{CJKutf8}
\usepackage{natbib}

\definecolor{lightblue}{RGB}{173,216,230}
\definecolor{lightgreen}{RGB}{200,230,201}
\definecolor{salmon}{RGB}{250,200,200}

\newcommand{\corraddress}{Meteorologisches Institut, Ludwig-Maximilians-Universität München, Theresienstraße 37, 80333 München, Germany}
\newcommand{\corremail}{Stefano.Franzoni@lmu.de}

\makeatletter
\renewcommand{\@maketitle}{%
  \newpage
  \begin{center}
    {\LARGE\bfseries \@title \par}
    \vskip 1em
    {\@author \par}
    \vskip 0.5em
    {\small \@date}
  \end{center}
  \vskip 1em
}
\makeatother

\begin{document}

\title{Using Distributional Regression Networks to Retrieve Cloud Properties from Solar Satellite Channels for Data Assimilation}

\author[1]{Stefano Franzoni}
\author[2]{Christopher Bülte}
\author[3]{Leonhard Scheck}
\author[4]{Christian Keil}
\author[5]{George C. Craig}

\affil[1]{Meteorologisches Institut, Ludwig-Maximilians-Universität München, Munich, Germany}
\affil[2]{Department of Mathematics, Ludwig-Maximilians-Universität München, Munich, Germany}
\affil[3]{Deutscher Wetterdienst, Munich, Germany}
\affil[4]{Meteorologisches Institut, Ludwig-Maximilians-Universität München, Munich, Germany}
\affil[5]{Meteorologisches Institut, Ludwig-Maximilians-Universität München, Munich, Germany}

\date{}
\maketitle

\noindent
\textbf{Correspondence}\\
Stefano Franzoni, \corraddress\\
Email: \href{mailto:\corremail}{\corremail}

\bigskip

%\linenumbers
\section*{Abstract}
Satellite observations in the solar part of the spectrum (including visible and near-infrared channels) offer high-resolution information on clouds and atmospheric properties that could be valuable for data assimilation. While forward operators suitable for the direct assimilation of these observations have become available recently and a first visible channel is already used operationally, their assimilation remains challenging due to strong non-linearities, ambiguities and high inter-channel correlations. This study addresses two central questions: what is the potential impact of assimilating multiple solar channels jointly, and can observed reflectances be transformed into physically meaningful, uncertainty-quantified variables better suited to assimilation than the raw reflectances themselves?
As a proof of concept, we assess the joint information content of six solar channels from the Flexible Combined Imager (FCI) onboard Meteosat Third Generation and introduce a novel "Backward Operator" (BO) for probabilistic retrievals of cloud-related variables. The BO is implemented in a machine learning approach as a distributional regression network that produces multivariate Gaussian estimates of total optical thickness, column cloud fraction, ice fraction, and effective radii of water and ice. The network is trained on synthetic images generated from regional ICON (Icosahedral Nonhydrostatic model) forecasts, and accounts for uncertainties in the input variables through a heteroscedastic noise model.
The BO provides unbiased and well-calibrated, situation-dependent predictions, with realistic covariance structures and reliable uncertainty estimates at the 1--$\sigma$ level. The retrieved variables can be overall usefully constrained, with a non-trivial uncertainty and covariance structure. Despite strong inter-channel correlations, combining multiple visible channels yields substantial performance improvements. As the BO does not require prior information, is fully consistent with an existing forward operator, and yields cloud variables that relate more linearly to the NWP model state, assimilating these variables could be a viable alternative to direct reflectance assimilation.

\section{Introduction}

Over the last decades, satellite images have played an increasingly central role as an observation source for data assimilation (DA) in operational global numerical weather prediction (NWP) systems. The spatio-temporal resolution of satellite imagers is higher than what observation networks can provide, and they also cover regions where observations are sparse or absent (e.g. over oceans). Currently, the combined satellite observations have the highest impact on the quality of forecasts of operational global NWPs (e.g., \citealt{bormann2019, eyre2022}), and have also become more central in regional models \citep[see e.g.][]{kurzrock2018}. Solar channels, commonly observing in the visible (VIS) and near-infrared (NIR) regions of the spectrum, have been available since the first weather satellites. In the current generation of imagers, with instruments like the Flexible Combined Imager (FCI, \citealt{Holmlund+2021}) on Meteosat Third Generation (MTG) or the AHI instrument \citep{Bessho+2016} on the Himawari satellites, there are similar numbers of solar and thermal channels, but the solar channels tend to have higher spatial resolutions. Therefore, the majority of data generated by these imagers stem from the solar channels, which provide valuable information about clouds, aerosols and trace gases.

Information on NWP model state variables can be inferred from these channels using retrieval methods (see, for example, \citealt{nakajima1990, wattsOCA2011, chen2022, gonzalez2025}). Cloud retrievals from solar channels have historically been used mainly for climate studies (e.g. \citealt{rossow1999}), model evaluation (e.g. \citealt{pincus2008}) and to understand cloud processes (e.g. \citealt{stephens2005}), rather than for data assimilation. Some retrievals based on satellite observations are assimilated routinely, with examples including 1D‑Var humidity and temperature profiles from infrared channels (e.g. \citealt{eyre1993assimilation}) and atmospheric motion vectors derived from solar channels \citep{bouttier2001}. As retrievals often rely on the information coming from a reference model (for example, in the Optimal Estimation framework, \citealt{rodgers2000}), the differences in modelling assumptions from those used in the DA system into which they are assimilated can introduce inconsistencies. Furthermore, when the model information comes from a closely related version of the system used to then assimilate the retrieval, background and retrieval errors may no longer be independent, posing additional challenges for ensuring well-calibrated error estimates in the DA cycle.

These problems can be avoided when satellite radiances are assimilated directly, which requires a forward operator to generate model equivalents from the NWP model state. Direct, all-sky assimilation has become the standard approach for microwave (MW) observations \citep{geer2018}. Also for infrared (IR) channels direct assimilation is the preferred way, and operational centres are in the process of moving from clear-sky to all-sky assimilation \citep{geer2019}. However, for solar channels, the situation is quite different. At the time of writing, only the $0.6\,\mu\mathrm{m}$ visible channel of SEVIRI (Spinning Enhanced Visible and Infrared Imager) onboard Meteosat Second Generation (MSG) is used operationally, in the regional data assimilation system of the German Weather Service (DWD). The direct assimilation of solar channels has long been hampered by the lack of suitable forward operators. However, MFASIS (Method for FAst Satellite Image Synthesis, see \citealt{scheck2016, baur2023}), distributed as part of the RTTOV package (Radiative Transfer for TOVS, see e.g. \citealt{saunders2018rttovv13}), supports now most solar channels, making their direct assimilation feasible.

With the increasing availability and space-time resolution of solar channels, and their support by MFASIS, the question arises of how they can best be exploited within DA systems. Although the direct assimilation of radiances is now generally the preferred way, it presents several challenges, in particular in ensemble-based DA frameworks like the Local Ensemble Transform Kalman Filter (LETKF) used in the regional DA system at DWD \citep{schraff2016}. First, the relationship between the model state and solar-channel radiances is strongly non-linear, in particular because the reflectance signal saturates at high optical depths. In combination with the linearity assumptions used in ensemble DA methods like the LETKF, this leads to a suboptimal analysis \citep{scheck2020, kugler2025}. Second, individual solar channels generally provide very ambiguous information, in the sense that many different model states can lead to the same signal. For instance, water and ice clouds tend to look similar in visible satellite images. Although using multiple solar channels would tend to resolve such ambiguities, the limited ensemble size of the DA system poses an upper limit to the number of observations that can be assimilated within a given localisation radius \citep{hotta2021}, meaning that including more satellite channels may result in an over-confident analysis. In addition, direct assimilation of several channels increases the computational cost. Third, the spatial and inter-channel correlation structure makes the assimilation of solar channels problematic. The impact of spatial correlations can be reduced by standard techniques like thinning and superobbing, at the cost of discarding part of the observations. However, solar channels also show high inter-channel correlations, with the visible channels being almost perfectly correlated. Properly accounting for inter-channel correlations would require a detailed specification of observation error covariances, and the additional information gained from assimilating further channels may therefore not justify the increased complexity \citep{TeresakiCorrelatedRMatrixLETKF}. These restrictions highlight the need to carefully assess what can actually be gained from directly assimilating solar channels. Ideally, estimates for the potential impact of multiple solar channels should not be influenced by current limitations of a certain DA system, but provide an independent assessment.

In this study, we present a retrieval method for the systematic quantification of the situation-dependent joint information content of multiple solar channels, providing useful insights into their potential observation impact. The method is also designed to yield retrievals that are suitable for assimilation, as it is compatible with the DA system (i.e., based on the same modelling assumptions and forward operator), does not rely on prior model information and provides calibrated, situation-dependent error covariance estimates. The proposed solution is based on a “Backward Operator” (BO), specified as a distributional regression network (DRN) that partially and approximately inverts the forward operator, providing probabilistic retrievals. We focus on the six solar MTG FCI channels supported by MFASIS in RTTOV 14.0, with a training dataset based on a Nature run from the regional ICON-D2 model over a 1-month summer period. The BO takes as inputs the set of 6 solar reflectances, their corresponding albedos, and the geometry of the scene, as specified by solar and viewing zenith angles, and by the scattering angle. The DRN then outputs a conditional distribution, parametrized as a multivariate Gaussian, for the natural logarithm of total optical thickness $\log(\tau)$, column cloud fraction $\gamma$, ice fraction $f_{\mathrm{ice}}$ and water and ice effective radii $r_{\mathrm{effw}}$, $r_{\mathrm{effi}}$. Uncertainty on the input variables can be accounted for during the training by injecting noise according to a heteroscedastic Gaussian noise model. The retrieved variables are more closely related to the model state compared to reflectances, and they also allow for physics-based characterization of the properties of the column. The proposed approach avoids the need of prior information, as the input-output relationship is learnt during training based on model climatology. Consistency between the BO and the NWP and DA system is ensured by using the same forward operator and NWP model adopted for the forward problem. Since the BO provides a probabilistic retrieval parametrized by a multivariate Gaussian, the output covariance matrix can be used as the error covariance matrix for the assimilation of retrieved quantities, once its representativeness component is properly accounted for following \cite{desroziers2005}. This approach therefore avoids typical drawbacks of standard retrieval methods, particularly in the context of DA, namely the reliance on prior information and the lack of reliable estimates of retrieval errors. In addition, the results of the proposed method are independent of the specific DA system adopted and are thus not affected by deficiencies or suboptimal performance of the chosen filter. Moreover, assimilating retrievals may offer additional advantages, as they typically exhibit lower correlations than reflectances in direct assimilation approaches, and they are more directly related to model state variables, with a potential to mitigate both ambiguity and nonlinearity errors. This study is intended as a proof of concept and, as such, involves limitations and simplifying assumptions, including the use of a regional domain and a climatology limited to a single summer month. The paper is organized as follows. In Section~\ref{section:data_methods}, the NWP model, the forward operator MFASIS and the training dataset are described (Sections~\ref{section:nwpmodel},~\ref{section:mfasis} and ~\ref{section:data}, respectively), and the DRN framework is presented (Section~\ref{section:DRNs}), with a description of the evaluation metrics in Section~\ref{sec:model_performance}. The results are then presented and discussed in Section~\ref{section:results}, followed by conclusions and outlook in Section~\ref{section:conclusion_outlook}.

\section{Data and Methods}\label{section:data_methods}

\subsection{Satellite Channels and their Sensitivities}\label{section:channels}

This study focuses on the solar channels of the Flexible Combined Imager (FCI) aboard the Meteosat Third Generation (MTG) geostationary platform. FCI includes eight solar channels that measure the intensity of sunlight reflected by Earth: five in the visible\footnote{Following EUMETSAT, we will refer to channels with wavelengths smaller than $1\,\mu$m as visible, even though the human eye is not sensitive to radiation with wavelengths larger than $0.7\,\mu$m.} (VIS), centred at $0.44\,\mu$m (VIS004), $0.51\,\mu$m (VIS005), $0.64\,\mu$m (VIS006), $0.87\,\mu$m (VIS008), and $0.91\,\mu$m (VIS009), and three in the near-infrared (NIR), centred at $1.38\,\mu$m (NIR014), $1.61\,\mu$m (NIR016), and $2.25\,\mu$m (NIR022). The VIS009 and NIR014 channels are strongly affected by water vapour absorption and will not be considered here, as our focus is on cloud properties. The remaining six channels are mainly sensitive to the optical thickness of water and ice clouds, as well as to the effective droplet and ice particle radii. The radius-sensitivity is significantly higher for the near-infrared channels, and the NIR016 channel is also sensitive to the cloud phase, with ice clouds leading to much lower reflectances than water clouds. For the channels with smaller wavelength, there is also some contribution from Rayleigh scattering, which is smaller than typical cloud signals and negligible for the near-infrared channels. In contrast to infrared channels, the cloud top height has only a very weak influence on the signal. Details of the vertical cloud structure are also not important. The vertically integrated cloud water and ice contents (or optical depths) and weighted vertical averages of the effective radii determine nearly all of the cloud signal in solar channels.

\begin{figure}[h]
    \centering
    \includegraphics[width=0.4\linewidth]{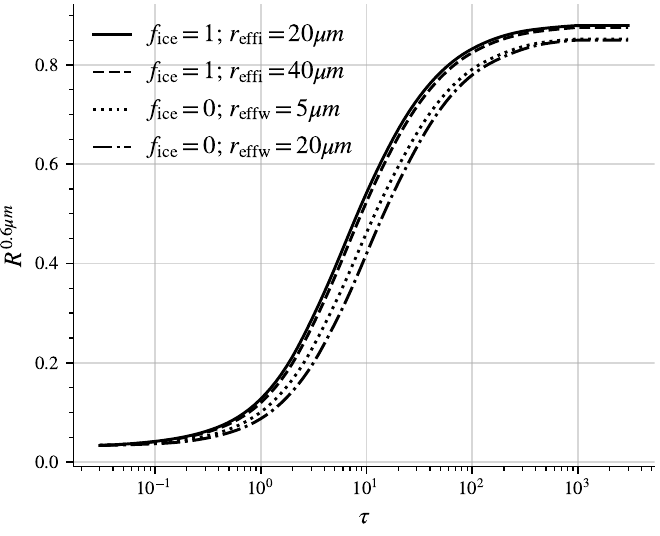}
    \caption{Reflectance of the 0.6$\mu$m FCI channel as a function of optical depth $\tau$ for different ice fractions $f_{\mathrm{ice}}$ (zero indicating a pure water cloud, one a pure ice cloud) and effective particle radii ($r_{\mathrm{effw}}$ for water, $r_{\mathrm{effi}}$ for ice clouds). }
    \label{fig:ambiguity}
\end{figure}

\begin{figure}
    \centering
    \includegraphics[width=0.85\linewidth]{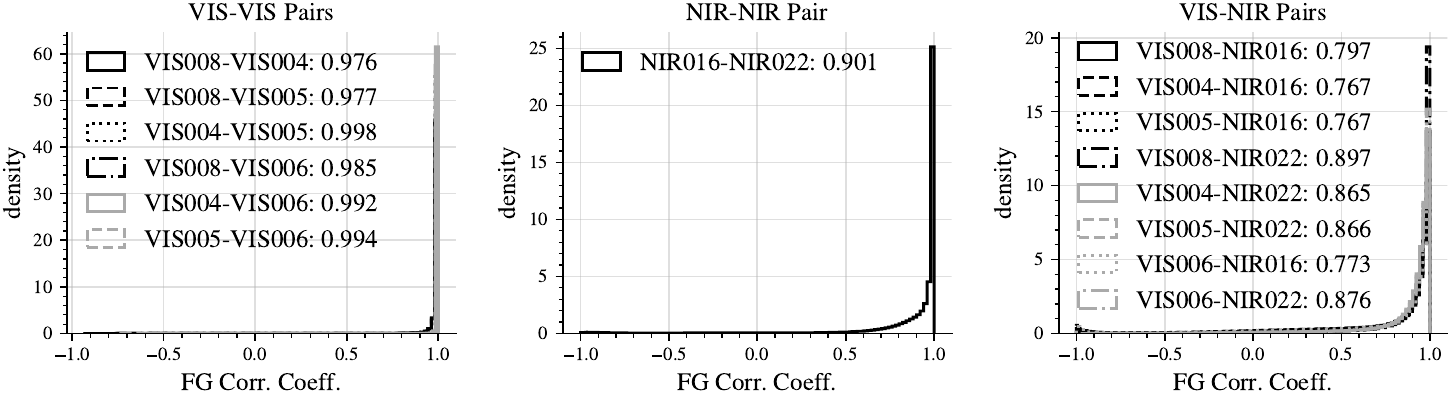}
    \caption{Histograms of the first guess reflectance ensemble correlation coefficients based on the 12UTC daily outputs of a ICON-D2 reference DA experiment with near-operational settings for the period between 16 August and 15 September 2022. In the legend, the averages of the absolute values of the first guess correlation coefficients are shown.}
    \label{fig:fg_correlation}
\end{figure}

Although these sensitivities of the solar channels to cloud properties represented in the model state are well known, considerable challenges for constraining the model state with direct reflectance assimilation exist. In particular, the ambiguity of individual channels and strong inter-channel correlations are problematic. Solar channels are ambiguous in the sense that many different cloud situations can lead to the same reflectance value. An example of ambiguity with respect to optical depth, effective radii, and cloud phase is given in Fig.~\ref{fig:ambiguity}. Unresolved clouds, characterized by a cloud fraction between zero and one in NWP models, present another source of ambiguity. A dense cloud occupying only part of a grid cell can lead to the same reflectance as a less dense cloud covering the full cell. The joint information coming from multiple solar channels may help in resolving these ambiguities. The interaction of photons with cloud droplets described by Mie scattering depends on the ratio of droplet size to wavelength, and therefore reflectance will vary slightly differently with effective radius for all of the channels. Further differences between the channels are related to Rayleigh scattering, surface albedo, and absorption in cloud particles, which all depend on the wavelength.
However, the joint impact of multiple channels on the analysis in a direct assimilation approach is affected and potentially limited by high inter-channel correlations. Correlation coefficients for synthetic FCI images computed for the 40 first guess members (at 12 UTC) of an ICON-D2 reference DA experiment with near-operational settings for the period between 16 August and 15 September 2022 show that the channels are indeed highly correlated (Fig.~\ref{fig:fg_correlation}). In particular, all pairs of visible channels have correlation coefficients with mean absolute values higher than $0.97$. The near-infrared channels are less correlated and the weakest correlations (approximately $0.8$ in the averaged absolute value) are found between NIR016 and any of the visible channels.

\subsection{NWP Model}\label{section:nwpmodel}

The training data set for this study is based on the daylight hourly output of a 30-day Nature run performed with the regional model ICON-D2 (ICOsahedral Non-hydrostatic, \citealt{zangl2015}), a convection-permitting model whose domain covers Germany and parts of its neighbouring countries. The run covers the period from 16 August to 15 September 2022, and is initialized with ICON-D2 analysis, with hourly boundary conditions provided by ICON-EU.

\subsection{Forward Operator}\label{section:mfasis}

To compute synthetic satellite images from ICON-D2 model states we use MFASIS (Method for FAst Satellite Image Synthesis, \citealt{scheck2016, baur2023}), a neural network-based forward operator for solar channels included in the RTTOV radiative transfer package \citep{saunders2018rttovv13}. We employ the version described in \cite{baur2023}, which was included in RTTOV v14 and utilizes a feed-forward neural network with 16 input nodes and eight hidden layers with 15 to 25 nodes per layer to generate reflectances. The definition of the input variables of the NN is based on the idea that the complex profiles of optical depth and effective radii generated by a NWP model can be replaced by a suitably chosen simplified profile without changing reflectance significantly. The parameters that describe the simplified profile are used as inputs for the NN. They include the ice and water optical depths and effective radii for a two-layer ice cloud and a two-layer mixed-phase cloud. Additional NN input variables are the vertically integrated water vapor content, surface and cloud top pressure, sun and satellite zenith angles, and the scattering angle. Finally, also the surface albedo is required to generate reflectances, and it is taken from an atlas included in RTTOV. The NN is trained using synthetic data generated with the discrete ordinate method included in RTTOV. Typically, MFASIS reflectance errors are smaller than 0.01 and inference takes less than a microsecond per sample on standard server CPUs.

In the computation of the MFASIS input variables, it is assumed that all grid cells are either cloud-free or fully cloud-covered. To account for partially cloudy grid cells with cloud fractions between zero and one, a subgrid cloud overlap scheme has to be employed. Here we use the "streams" method \citep{matricardi2005} included in RTTOV, which assumes that clouds in adjacent layers of a model column overlap maximally and clouds separated by cloud-free layers overlap randomly. The streams method generates $n$ subcolumns with weights $w_i$, in which clouds are placed according to these random-maximum overlap rules. MFASIS is then called for each subcolumn and generates a reflectance $R_i$. Unless at least one of the grid cells in the column is fully cloudy, there will also be one cloud-free subcolumn, for which MFASIS is also called and generates the clear-sky reflectance $R_{clr}$. The reflectance for the full model column is simply $R=\sum_{i} R^\lambda_i w_i$. A by-product of this process is the column cloud fraction $\gamma \in [0,1]$, defined as one minus the weight of the cloud-free subcolumn. With this definition, the reflectance for the full column can also be written as $R=(1-\gamma)R_{clr}+\gamma R_{cloudy}$, where $R_{cloudy}$ is the weighted average reflectance of the cloudy part of the column. In other words, the column cloud fraction represents the fraction of the column in which a cloud is present in at least one of the layers.

\subsection{Training Data for the Backward Operator}\label{section:data}

Input and target variables for the Backward Operator are obtained by running the forward operator  (see~\ref{section:mfasis}) on the ICON-D2 vertical profiles from the Nature run described in~\ref{section:nwpmodel}. The input variables are reflectances and albedos for the six channels VIS004, VIS005, VIS006, VIS008, NIR016 and NIR022, the sun and satellite zenith angles and the scattering angle. The target variables are computed from the cloud input variables of the MFASIS NN. There are ten variables describing cloud optical depths and effective radii for the two-layer ice and mixed-phase clouds in MFASIS \citep[for details see][]{baur2023}. To reduce complexity and have a better chance to constrain important cloud properties with only six channels, we reduce these ten variables to four: the natural logarithm of total optical depth $\log(\tau)$, the ice fraction $f_{\mathrm{ice}}$, defined as the ratio of the total ice optical depth to $\tau$, and the mean effective radii for water and for ice clouds, $r_{\mathrm{effw}}$ and $r_{\mathrm{effi}}$, respectively. In the forward operator, these cloud variables are computed for each subcolumn generated by the cloud overlap scheme. As target variables for the BO we use average values of the four variables for the cloudy part of the column, and the column cloud fraction $\gamma$ that quantifies the size of the cloudy part. Since these target variables are derived directly from the Nature run, they are treated as ground truth throughout this work, and the two terms are used interchangeably hereafter.

\begin{figure}
    \centering
    \includegraphics[width=0.75\linewidth]{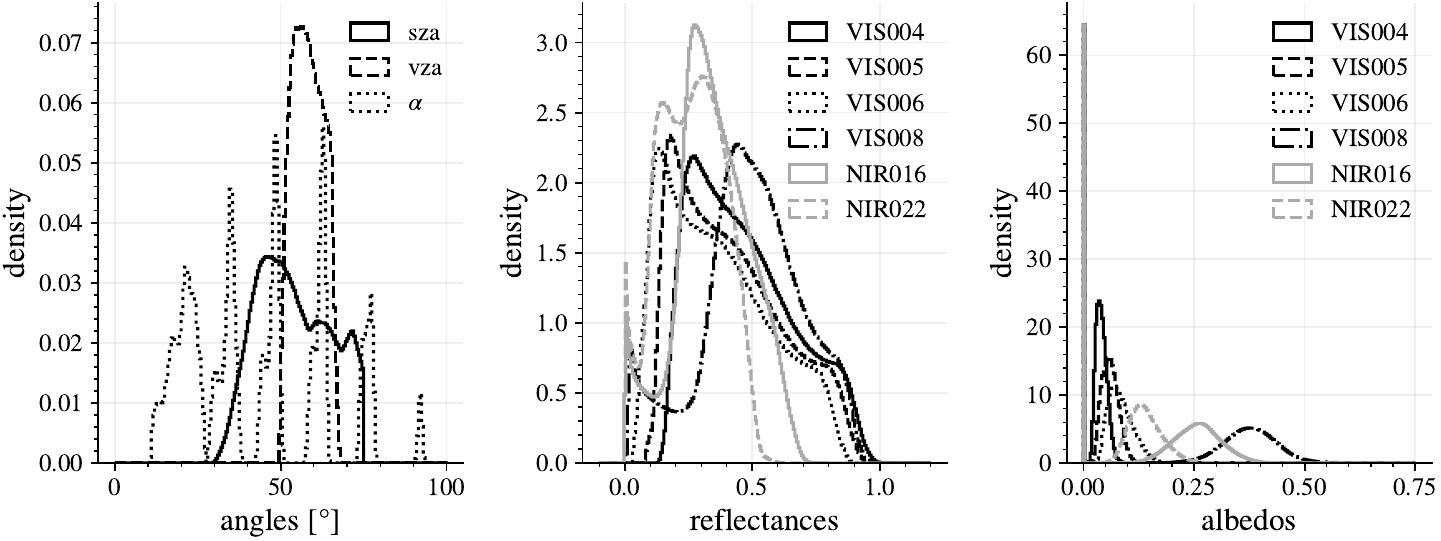}
    \caption{Training dataset histograms of the NN input variables, including angles (solar zenith angle sza, viewing zenith angle vza and scattering angle $\alpha$), reflectances and albedos for the set of MTG solar channels considered here (namely, VIS004, VIS005, VIS006, VIS008, NIR016, NIR022). The training dataset contains $\approx25\times10^6$ points, i.e. 80\% of the full dataset.}
    \label{fig:training_input}
\end{figure}

\begin{figure}
    \centering
    \includegraphics[width=0.75\linewidth]{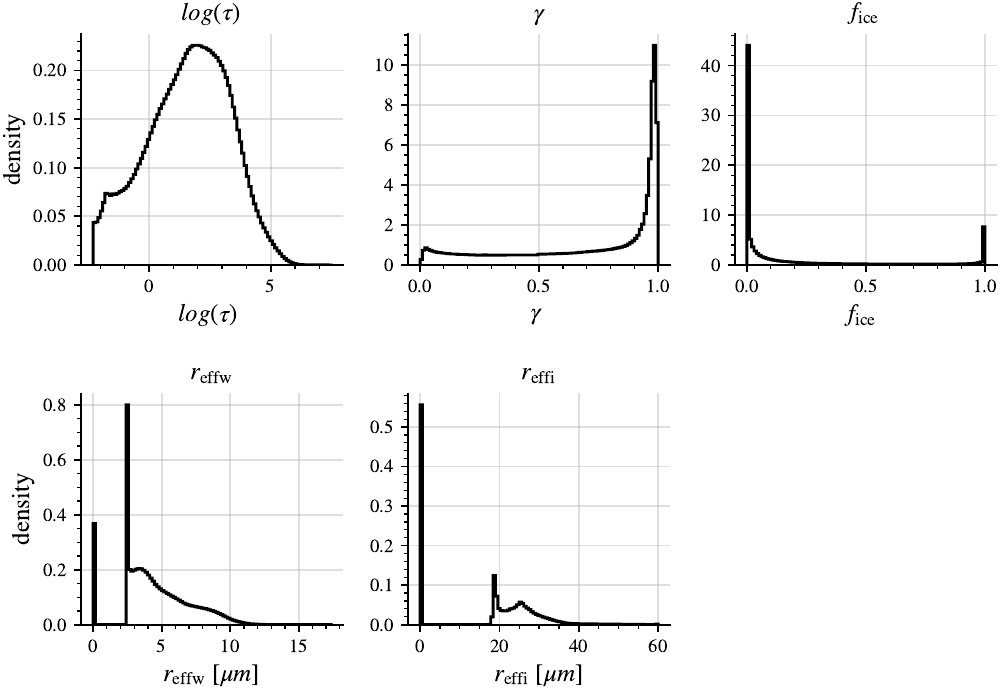}
    \caption{Training dataset histograms of the NN output variables, including the natural logarithm of total optical thickness $\log(\tau)$, column cloud fraction $\gamma$, ice fraction $f_{\mathrm{ice}}$ and water and ice effective radii $r_{\mathrm{effw}}$ and $r_{\mathrm{effi}}$, respectively. The training dataset contains $\approx25\times10^6$ points, i.e., 80\% of the full dataset.}
    \label{fig:training_output}
\end{figure}

The set of satellite pixels is then filtered to remove points with solar zenith angles higher than $75\degree$ (where 3D effects become relevant and the forward operator provides less realistic results), and with total optical thickness smaller than 0.1 (where the cloud signal is weak), resulting in a total of $\approx31\times10^6$ points. The data set is then randomly split into training (80\%), validation (10\%) and testing (10\%) subsets. Figure~\ref{fig:training_input} shows histograms of the input training variables. For simplicity, MFASIS is set up assuming zero albedo over the resolved bodies of water (i.e. oceans and lakes), which results in the peaks at zero in the albedo histograms. In Fig.~\ref{fig:training_output}, histograms for the target variables are shown for the training dataset. A wide variety of cloudy situations is represented, with clouds as thick as $\tau\approx 150$, including also cases where the column is not fully occupied by a cloud (i.e., $\gamma$ strictly in between 0 and 1). By convention, water and ice effective radii are set to 0 when ice fraction is 1 and 0, respectively.

\subsection{Modeling with Distributional Regression Networks}\label{section:DRNs}

As introduced above, we aim to learn a backward operator that approximately inverts the forward operator and predicts the parameters of interest $\log(\tau), \gamma, f_{\mathrm{ice}}, r_{\mathrm{effw}}, r_{\mathrm{effi}}$. However, the mapping from the input variables (including reflectances, albedos and the geometric angles) to these parameters is generally ill-posed, meaning that multiple solutions might exist for the input state, as is typical in inverse problems. A conditional probability distribution is therefore a natural choice to represent such a mapping, as the potentially multiple solutions contribute to the variance of the distribution. An additional source of variance arises from accounting for instrumental noise on the input reflectances and retrieval errors on the surface albedos, as will be discussed in Section~\ref{sec:noise_model}. For these reasons, we propose to learn the full predictive distribution of the output variables, conditioned on the input, using distributional regression networks (DRNs, \citealt{rasp2018}). DRNs have demonstrated strong performance in forecasting and probabilistic modeling and have been applied extensively to meteorological prediction tasks \citep{rasp2018,schulz2022, chapman2022}.

Assume that we have a finite dataset of input-target-pairs $\mathcal{D} = \{(x_i, y_i)_{i=1}^n\}$ with input features $x_i \in \mathbb{R}^k$ and target features (or response variables) $y_i \in \mathbb{R}^d, \ i =1, \ldots, n$. Let $P(\cdot \mid x_i)$ denote the true conditional probability distribution and $p(\cdot \mid x_i)$ the corresponding conditional probability density for a given observation $i$. It is worth underlining that, in this section, following the conventional meaning in this context, the term observation refers to the verifying outcome against which the predictive distribution is evaluated. In the distributional regression framework, we assume that the conditional distribution can be expressed in a parametric form as
\begin{equation}
    p(y_i \mid x_i) = p(y_i \mid \theta(x_i)),
\end{equation}
where $\theta(x_i) = (\theta_1(x_i), \ldots, \theta_p(x_i))^\top \in \Theta \subseteq \mathbb{R}^p$ denotes the covariate-dependent parameter vector \citep{KNEIB202399}. Each distributional parameter $\theta_j, \ j=1, \ldots, p$ is then a function of the covariates $x_i$, specific to the individual observation $i$. In particular, this link is established using parameter-specific regression predictors $\eta_j: \mathbb{R}^k \to \mathbb{R}$ via
\begin{equation}
    \theta_j(x_i) = h_j(\eta_j(x_i)), \quad j=1, \ldots, p.
\end{equation}
Here, $h_j: \mathbb{R} \to \mathbb{R}$ is the link function, mapping the predictor values to the support of each parameter, for example, the exponential function to ensure nonnegativity. The formulation above includes many commonly used models, such as linear and generalized additive models \citep{KNEIB202399} or ensemble model output statistics (EMOS, \citealt{gneiting2005}), but also neural network-based approaches. The latter, also known as distributional regression networks \citep{rasp2018} arises when the predictor for the parameter $\theta_j$, $\eta_j(x_i) = f_\phi^{\theta_j}(x_i)$ is parametrized with a neural network $f_\phi^{\theta_j}$ with weights (and biases) $\phi$.

\subsubsection{Network Specification and Probabilistic Model}
Distributional regression networks typically employ a predictive distribution that is either univariate and parametric \citep{rasp2018, schulz2022} or multivariate and non-parametric \citep{pacchiardi}. However, in our setting, we are mainly interested in modeling the first two moments of the distribution, i.e., the mean vector $\mu \in \mathbb{R}^d$ and the covariance matrix $\Sigma \in \mathbb{R}^{d\times d}$. In particular, we are interested in the marginal variances, as they represent the uncertainty per variable, as well as the predicted covariance structure, as it allows us to model the dependency structure of the different parameters. Consequently, both univariate and nonparametric methods are not applicable; the univariate, as it does not model covariance structure, and the nonparametric, because it would require an additional estimator of the covariance matrix, creating a computational overhead.

As a predictive model, we therefore employ a multivariate Gaussian distribution, which is only defined via its first two moments and allows for simultaneous assessment of marginal uncertainty, as well as the correlation structure of the output variables. While a Gaussian might seem restricting, in section~\ref{sec:gaussianity} we show that this assumption is justifiable for the majority of the considered variables. Under this parametrization, we aim to learn the following predictive density
\begin{equation}
    p(\cdot \mid \theta(x)) = \mathcal{N}(\cdot; \mu(x), \Sigma(x)),
\end{equation}
with input-dependent mean vector $\mu(x) \in \mathbb{R}^d$ and covariance matrix $\Sigma(x) \in \mathbb{R}^{d \times d}$. However, directly parametrizing $\Sigma$ with a neural network is not straightforward, as one needs to ensure that $\Sigma$ is symmetric and positive definite. To achieve this, we follow \cite{muschinski}, and consider a learned Cholesky decomposition of the covariance matrix $\Sigma = L L^\top$, where
\begin{equation}
\label{eq:cholesky_decomposition}
L^\top =\begin{pmatrix}
\lambda_{11} & \lambda_{12} & \cdots & \lambda_ {1d}\\
0 & \lambda_{22} & \cdots & \lambda_{2d}\\
\vdots & \vdots & \ddots & \vdots \\
0 &0 & 0 & \lambda_{dd}
\end{pmatrix} \in \mathbb{R}^{d \times d},
\end{equation}
is an upper triangular matrix with $d$ positive diagonal elements $\lambda_{ii} > 0$ and $d(d-1)/2$ off-diagonal elements. In particular, this decomposition is unique for a positive diagonal, and we can employ a Softplus activation in the last layer of the neural network as a link function $h_{\lambda_{ii}}: \mathbb{R} \to \mathbb{R}_+$, leading to the following parametrization:
\begin{align}
    \mu &= f^\mu_\varphi(x), \\
    \lambda_{ii} &= \mathrm{Softplus}\left(f^{\lambda_{ii}}_\varphi(x) \right), \quad i=1,\ldots,d, \\
    \lambda_{ij} & = f^{\lambda_{ij}}_\varphi(x) , \quad i=1,\ldots,d-1, \ j = i+1, \ldots, d, 
\end{align}
where $f^{\theta_j}_\varphi(x)$ denotes the neural network prediction for the parameter $\theta_j$.

\subsubsection{Proper Scoring Rules}\label{sec:scoring_rules}

To optimize the neural network with respect to the optimal predictive distribution, a suitable loss function is required. To this end, we make use of proper scoring rules \citep{gneiting2007scoringrules}, which have been extensively used in forecast evaluation and optimization of distributional regression networks.
A scoring rule $S$ assigns a numerical score to a predictive distribution $F$, in our case $F = \mathcal{N}(\mu, \Sigma)$, and a corresponding realized observation $y$. $S$ is called proper if the true distribution of the observation receives the minimal score in expectation, i.e.
\begin{equation}
\label{eq:sr_definition}
    \mathbb{E}_{Y \sim G}[S(G,Y)] \leq \mathbb{E}_{Y \sim G}[S(F,Y)], \quad \forall F,G \in \mathcal{P},
\end{equation}
where $\mathcal{P}$ denotes a suitable class of distributions (see \citealt{gneiting2007scoringrules} for details). Furthermore, $S$ is called strictly proper if \eqref{eq:sr_definition} holds with equality if and only if $F=G$. This general framework includes various commonly used loss functions, such as the negative log-likelihood or log score (LogS), the (mean) squared error (SE), or the energy score, which is a multivariate version of the commonly used continuous ranked probability score (CRPS, \citealt{gneiting2007scoringrules}). Closed-form expressions for the different scores and a multivariate predictive Gaussian can be found in Table~\ref{tab:scoring_rules}.
\begin{table}
\centering
\begin{adjustbox}{max width=\textwidth}
    \begin{tabular}{l|cc}
\toprule
& $S(\mathcal{N}(\mu, \Sigma), y)$ & Propriety\\
\midrule
Squared error (SE) &  $\| y - \mu \|_2^2$ & Proper\\
Log score (LogS) & $\frac{d}{2}\log(2\pi) + \frac{1}{2}\log(\det \Sigma) + \frac{1}{2}(y-\mu)^\top \Sigma^{-1}(y-\mu)$ & Strictly proper \\
Energy score (ES) & $ \mathbb{E}_{X \sim \mathcal{N}(\mu, \Sigma)}[\| X -y\|_2] - \frac{1}{2}\mathbb{E}_{X, X' \sim \mathcal{N}(\mu, \Sigma)}[\| X -X'\|_2]$ & Strictly proper \\
\bottomrule
\end{tabular}

\end{adjustbox}
\caption{Comparison of different scoring rules, their propriety and their analytical expressions for a predictive multivariate Gaussian.}
\label{tab:scoring_rules}
\end{table}

The (mean) squared error is the most commonly used loss function for regression tasks, but is insufficient for probabilistic models, since it only evaluates the mean prediction and ignores any higher-order moments and therefore is not strictly proper. The log-score (or negative log-likelihood) is a commonly used scoring rule and admits a closed-form expression for a multivariate Gaussian, as shown in Table~\ref{tab:scoring_rules}. However, the log score is a local scoring rule: it only evaluates the predicted distribution at the realized outcome and has been criticized for leading to uninformative and less calibrated predictions \citep{buchweitz2025asymmetricpenaltiesunderlieproper}. In addition, the score can diverge for small predicted variances, making optimization numerically unstable \citep{NEURIPS2019_07211688, NEURIPS2023_a901d554}.
Therefore, in this article, we focus mainly on the energy score, which has been extensively utilized in the training of neural network-based predictive models \citep{pacchiardi, engression}, with applications in fields such as meteorology \citep{rasp2018, chapman2022, chen_generative}, energy systems \citep{MAYER2026114361,chen2025probabilisticintradayelectricityprice}, or dynamical systems \citep{bultepno}. While the energy score does not admit a closed-form solution for a multivariate Gaussian \citep{szekely_test}, it can be efficiently approximated by an unbiased Monte Carlo estimator \citep{pacchiardi}. While our main results are obtained by training the energy score, we also provide comparisons to the squared error and the log score.

\subsubsection{Modelling Errors of the Input Variables}\label{sec:noise_model}

For practical applications of the method presented here, accounting for instrumental error on reflectances and retrieval uncertainty on albedos is crucial in order to get a realistic probabilistic prediction. We model such sources of uncertainty as an additive, uncorrelated Gaussian noise with zero mean and prescribed, heteroscedastic standard deviation. The model for the noise standard deviation for reflectances is inspired by the information on radiometric noise and calibration errors reported in the Flexible Combined Imager (FCI) Level 1 Product Validation Report \citep{FCI_val_report}. We define the noise standard deviation $\sigma_R^\lambda$ for the reflectance $R_\lambda$ for channel $\lambda$ as:
\begin{equation}\label{eq:noise_refl}
    \sigma_R^\lambda = \text{NEdL}_\lambda\frac{\pi}{I_\lambda\cos(\text{sza})}+0.03R_\lambda
\end{equation}
where $\text{NEdL}_\lambda$ is the channel radiometric noise, $I_\lambda$ the channel solar irradiance, and sza is the solar zenith angle. The two terms in the sum represent the contributions from radiometric noise (first term) and the one from calibration error (second term). The albedos noise standard deviation $\sigma_A^\lambda$ is defined in a piecewise way as follows:
\begin{equation}\label{eq:noise_albedo}
    \sigma_A^\lambda = \min\{A_\lambda/3,\,0.04\}
\end{equation}
where $A_\lambda$ is the surface albedo of the channel $\lambda$. The constant 0.04 value is motivated by information on the Bidirectional Reflectance Distribution Function (BRDF) atlas presented in \cite{BRDFatlas}, whereas a constant relative uncertainty of 30\% is assumed for very low albedos (e.g., over bodies of water). This albedo error model is meant as a simple starting point for this proof of concept study and neglects e.g. the contribution of sun glint to the ocean albedo, which depends on near surface wind speed \citep[see e.g.][]{saunders2018rttovv13}.

\subsection{Metrics for assessing the Models Performance}\label{sec:model_performance}

The general approach to assess the DRN's performance is based on evaluating the quality of the predicted conditional distribution, and on testing a-posteriori the validity of the modelling assumptions. The evaluation is performed on the test dataset (see Sec.\ref{section:data}), which includes $\approx3.1\times 10^6$ points. In the multivariate Gaussian case assumed here, mean and covariance predictions are assessed, both separately and jointly. Mean predictions are tested with standard metrics like bias, Root Mean Squared Error (RMSE) and Pearson R coefficient. In addition, we compare predicted means and covariances with the sample estimates from the method described in \ref{section:sampling_covariances}. The predicted distribution is tested for calibration, in the spirit of \cite{gneiting2005, gneiting2007calibrationsharpness}. The motivation and definitions of the metrics is provided in the following sections.

\subsubsection{Predicted Mean}

The predicted means are tested with standard metrics, namely, bias, Root Mean Squared Error (RMSE) and Pearson R coefficient. Letting $\mu_i$ be the predicted mean of one of the output variables for the $i$-th point in the test dataset, $y_i$ the associated true value, and $N$ the size of the test dataset, the bias can be defined as $\text{bias}_i=\mu_i -y_i$, whereas RMSE and Pearson R as:
\begin{equation*}
    \text{RMSE}=\sqrt{\frac{1}{N}\sum_i (\mu_i -y_i)^2}
    \hspace{15mm}\mathrm{R}\left[\mu,y\right]=\frac{\sum_i (\mu_i -\overline{\mu})(y_i-\overline{y})}{\sqrt{\sum_i(\mu_i -\overline{\mu})^2}\sqrt{\sum_i(y_i-\overline{y})^2}}
\end{equation*}
where $\overline{\mu}$  and $\overline{y}$ denote the arithmetic average of the predicted means and of the true values, respectively, taken over the test dataset. Using the sampled conditional distribution described in Sec.\ref{section:sampling_covariances}, the Pearson R coefficient can be computed also between sampled means, predicted means and truth. As the NN is trained to provide the best estimate of a Gaussian parametrization of the conditional distribution, sampled and predicted means are expected to show the highest correlation compared to the other pairings.

\subsubsection{Predicted Conditional Distribution}

Given the probabilistic character of the NN predictions, it is key to assess the quality of the predicted Gaussian conditional distribution and the validity of the underlying Gaussianity assumption. As suggested in \cite{gneiting2007calibrationsharpness}, we evaluate the calibration and sharpness of the predictive distribution jointly using a proper scoring rule. Calibration is a joint property of the probabilistic prediction and the ground truth, referring to the statistical consistency between the two. Sharpness refers to the predictive distribution's capacity to effectively constrain the predicted variables. Proper scoring rules assess calibration and sharpness jointly, as they reach a minimum in the case of an ideal prediction. We rank the model versions considered in this study using the energy score used in the training phase, as it is suitable for our multivariate setting.

The best performing model is then further analysed considering calibration in a univariate sense. Taking marginals of the multivariate Gaussian distribution is straightforward, leading to a set of independent Gaussian distributions parametrized by the corresponding mean and standard deviation. For each NN output variable, calibration is assessed with the histogram of the Probability Integral Transform (PIT). Denoting with $F_C$ the cumulative distribution corresponding to the predictive distribution $F$ and with $y$ the ground truth, PIT is simply defined as $p=F_C(y)$. A standard proof shows that, if the predictive distribution is perfectly calibrated and $F_C$ is continuous, then the PIT follows a uniform distribution $U[0,1]$ between 0 and 1. Hence, PIT histograms can be directly compared to the $U[0,1]$ as a test for calibration. It must be noted though that uniformity is only a necessary condition for calibration. Histograms of PIT are also known to enhance departures from uniformity with large sample sizes, so that 10 to 20 bins is a generally good binning strategy \citep{gneiting2007calibrationsharpness}. Finally, we also consider coverage as a metric to test calibration at a given confidence level. Following \cite{neyman1937}, coverage can be defined here operatively as the fraction of ground truth points $y_i$ in the test dataset contained in a confidence interval centred around the predicted mean $\mu_i$. The nominal coverage can then be evaluated in the hypothesis that the predicted distribution is exact. Assuming Gaussianity, the half-width of the interval is customarily expressed as integer multiples of the predicted standard deviation $\sigma$. Here we focus on the 1--$\sigma$ coverage, with well-known nominal value of $\approx0.68$, as it aligns with the anticipated applications of the method. Indeed, if used purely as a retrieval method, the predicted sigma are taken as a proxy for the error, whereas, in a DA context, the most common available filters rely on Gaussianity as an effective modelling assumption. Thus, both applications make the estimate of the standard deviation particularly important.

\subsubsection{Sample Estimate of the Predicted Conditional Distribution}\label{section:sampling_covariances}

In this subsection, we describe a method to sample the conditional distribution predicted by the NN from the training data. Analogously to the NN training and test phases, the source of information to sample the conditional distribution is represented by the training dataset, whereas the test dataset provides reference points to validate the results. By construction, the NN is trained to predict a parametrized version of the conditional probability density $p(y \mid x)$ with $x\in\mathbb{R}^k$ and $y\in\mathbb{R}^d$ being the input and output space vectors. Reflectances and albedos in $x$ are assumed to be Gaussian distributed with standard deviation given by error models. The angles (namely, solar and viewing zenith angles and scattering angle) are assumed to be perfectly known. With this at hand, it is possible to sample $p(y \mid x)$ from the training data. Let us select a set of $N_{points}$ reference points $\{x_i\}_i$ from the test dataset. For each $x_i$, the likelihood of observing any point in the training dataset can be computed in the input space, taking as mean the selected reference point $x_i$, and as standard deviation the observation errors provided by the error model for reflectances and albedos in Equations~\ref{eq:noise_refl} and ~\ref{eq:noise_albedo}. In order to ensure statistical convergence of the sampling method, the angles in the input vector are also assumed to be independently, gaussianly distributed, with a $0.5\degree$ uncertainty. Then, the likelihoods over the training dataset are used as weights to sample $N_{sample}$ points from the training dataset. Considering the set of output variables corresponding to the sampled points, such set samples the sought-after distribution $p(y \mid x)$. The output mean and covariance matrix can then be estimated as the sample mean and the sample covariance matrix, respectively. By repeating the procedure for all the reference points, statistics of the means, covariances and standard deviations can be obtained, which can be compared to those predicted by the NN and, where possible, to the truth. For this work, $N_{points}=10^4$ and $N_{sample}=10^2$.
The method described above can be seen as a Monte Carlo sampling of the estimator for the conditional probability density function in \cite{hyndman1996}. Specifically, the bandwidth for $x$ is physically motivated for reflectances and albedos and set to the observation errors, while it is set to $0.5\degree$ for the input angles. A zero bandwidth is chosen for $y$, where no smoothing is assumed, so that estimators for the mean and variance are asymptotically unbiased and correspond to the arithmetic mean and sample variance, respectively. A Monte Carlo version was preferred to improve numerical stability and computational cost, considering the size of the training dataset.

\section{Results}\label{section:results}

The six models we consider differ by the choice of loss function and by whether input reflectances and albedos are treated as noisy or not. As discussed in~\ref{sec:scoring_rules}, we consider three loss functions, namely, squared error, energy score, and log score. For each of these models, two versions are trained, one including noisy input reflectances and albedos, the other assuming them to be known exactly. The noise model is specified in Section~\ref{sec:noise_model}, and the noise standard deviations are defined by Equations~\ref{eq:noise_refl} and ~\ref{eq:noise_albedo}. The input features consist of six solar MTG reflectance channels (namely, VIS004, VIS005, VIS006, VIS008, NIR016, NIR022), the corresponding albedos, the solar and viewing zenith angles, and the scattering angle. The output variables comprise the natural logarithm of the total optical thickness of the VIS006 channel\footnote{Optical depth is channel specific, but can be converted to channel-independent quantities like the total column cloud water and ice contents.} $\log(\tau)$, column cloud fraction $\gamma$, ice fraction $f_{\mathrm{ice}}$ and water and ice effective radii $r_{\mathrm{effw}}$, $r_{\mathrm{effi}}$. A summary of the model features and naming conventions is given in Table~\ref{tab:models}. An additional nine models are described in Section~\ref{sec:sensitivity}, which are designed to test the sensitivity of retrieval performance to the number and type of input channels. 
\begin{table}
    \centering
    \begin{adjustbox}{max width=\textwidth}
        \begin{tabular}{l|cccccc}
    \toprule
    & Det-Noise-RMSE & MVG-Noise-Energy & MVG-Noise-Log & Det-RMSE & MVG-Energy & MVG-Log \\
    \midrule
    Distribution & Deterministic & MVG & MVG & Deterministic & MVG & MVG \\
    Loss & SE & ES & LS & SE & ES & LS \\
    Noise & On & On & On & Off & Off & Off \\
    \bottomrule
\end{tabular}
    \end{adjustbox}
    \caption{Summary of the models considered in this study, showing the assumed distribution, the training loss and the use of noisy input reflectances and albedos. Here, MVG is a shorthand for Multivariate Gaussian distribution, SE for Mean Squared Error, ES for Energy Score and LS for Log Score.}
    \label{tab:models}
\end{table}

\subsection{Global Metrics} %  for All Models
In Table~\ref{tab:global_metrics}, a performance comparison between the model versions listed in Tab.~\ref{tab:models} is presented, based on Energy Score (ES), RMSE, Pearson R and bias (see Section~\ref{sec:model_performance}). Log-score-based models MVG-Noise-Log and MVG-Log showed instabilities during training due to small predicted variances and their local scoring rule, as discussed in Section~\ref {sec:scoring_rules}. This is reflected in a higher ES and RMSE compared to their ES-based counterparts. Comparing Det-Noise-RMSE and MVG-Noise-Energy, the probabilistic approach can be seen to be beneficial for a lower RMSE (5-8\% reduction in the noisy version). Similarly, the no-noise version MVG-Energy tends to outperform the deterministic, no-noise Det-RMSE one, even though by a small margin. The behaviour outlined above is expected, as the probabilistic approach is more suitable and better in dealing with noisy data, while converging to the deterministic version when data have no noise. Overall, as expected, noise injection increases the ES and RMSE compared to zero-noise versions. Nevertheless, with the chosen noise models, the impact is found to be relatively limited when compared to the zero-noise versions, making the predictions useful also in the noisy case. Looking at the Pearson R between mean predictions and ground truth, the same patterns can be identified. The mean bias also behaves similarly, with higher bias (in absolute value) for the Log-score-based models, though all models exhibit very low values. Given the considerations above, we regard the ES-based models as the best performing versions, focusing the analysis on the MVG-Noise-Energy model, as it is more relevant for practical applications. The MVG-Energy version will be mainly used to showcase the differences in predicted uncertainties between the noise and no-noise versions.

\begin{table}
\centering
\begin{adjustbox}{max width=\textwidth}
    \begin{tabular}{lrrrrrr}
\toprule
 & Det-Noise-RMSE & MVG-Noise-Energy & MVG-Noise-Log & Det-RMSE & MVG-Energy & MVG-Log \\
\midrule
ES & & & & & & \\
\midrule
All variables & - & {\cellcolor{lightblue!49}} 3.962 & {\cellcolor{lightblue!100}} 6.081 & - &{\cellcolor{lightblue!15}} 2.520 & {\cellcolor{lightblue!74}} 5.010 \\
\midrule
RMSE & & & & & & \\
\midrule
$\log(\tau)$  & {\cellcolor{lightblue!68}} 0.5658 & {\cellcolor{lightblue!60}} 0.5403 & {\cellcolor{lightblue!100}} 0.6640 & {\cellcolor{lightblue!19}} 0.4093 & {\cellcolor{lightblue!15}} 0.3964 & {\cellcolor{lightblue!59}} 0.5352 \\
$\gamma$  & {\cellcolor{lightblue!73}} 0.1647 & {\cellcolor{lightblue!62}} 0.1517 & {\cellcolor{lightblue!100}} 0.1937 & {\cellcolor{lightblue!16}} 0.1013 & {\cellcolor{lightblue!15}} 0.09967 & {\cellcolor{lightblue!59}} 0.1486 \\
$f_{\text{ice}}$  & {\cellcolor{lightblue!67}} 0.1818 & {\cellcolor{lightblue!58}} 0.1670 & {\cellcolor{lightblue!100}} 0.2371 & {\cellcolor{lightblue!15}} 0.09425 & {\cellcolor{lightblue!15}} 0.09353 & {\cellcolor{lightblue!49}} 0.1518 \\
$r_{\text{effw}}$ [$\mu m$] & {\cellcolor{lightblue!79}} 1.506 & {\cellcolor{lightblue!67}} 1.412 & {\cellcolor{lightblue!100}} 1.669 & {\cellcolor{lightblue!18}} 1.023 & {\cellcolor{lightblue!15}} 0.9926 & {\cellcolor{lightblue!69}} 1.424 \\
$r_{\text{effi}}$ [$\mu m$] & {\cellcolor{lightblue!71}} 9.315 & {\cellcolor{lightblue!53}} 8.401 & {\cellcolor{lightblue!100}} 10.74 & {\cellcolor{lightblue!17}} 6.604 & {\cellcolor{lightblue!15}} 6.490 & {\cellcolor{lightblue!73}} 9.401 \\
\midrule
Pearson R & & & & & & \\
\midrule
$\log(\tau)$  & {\cellcolor{lightblue!51}} 0.9420 & {\cellcolor{lightblue!59}} 0.9472 & {\cellcolor{lightblue!15}} 0.9194 & {\cellcolor{lightblue!96}} 0.9701 & {\cellcolor{lightblue!100}} 0.9719 & {\cellcolor{lightblue!62}} 0.9487 \\
$\gamma$  & {\cellcolor{lightblue!48}} 0.8543 & {\cellcolor{lightblue!61}} 0.8787 & {\cellcolor{lightblue!15}} 0.7936 & {\cellcolor{lightblue!98}} 0.9475 & {\cellcolor{lightblue!100}} 0.9494 & {\cellcolor{lightblue!64}} 0.8835 \\
$f_{\text{ice}}$  & {\cellcolor{lightblue!59}} 0.8366 & {\cellcolor{lightblue!69}} 0.8656 & {\cellcolor{lightblue!15}} 0.7006 & {\cellcolor{lightblue!99}} 0.9588 & {\cellcolor{lightblue!100}} 0.9596 & {\cellcolor{lightblue!76}} 0.8894 \\
$r_{\text{effw}}$ [$\mu m$] & {\cellcolor{lightblue!40}} 0.7764 & {\cellcolor{lightblue!54}} 0.8073 & {\cellcolor{lightblue!15}} 0.7191 & {\cellcolor{lightblue!97}} 0.9038 & {\cellcolor{lightblue!100}} 0.9099 & {\cellcolor{lightblue!54}} 0.8088 \\
$r_{\text{effi}}$ [$\mu m$] & {\cellcolor{lightblue!51}} 0.7100 & {\cellcolor{lightblue!71}} 0.7755 & {\cellcolor{lightblue!15}} 0.5850 & {\cellcolor{lightblue!97}} 0.8665 & {\cellcolor{lightblue!100}} 0.8738 & {\cellcolor{lightblue!50}} 0.7059 \\
\midrule
Bias & & & & & & \\
\midrule
$\log(\tau)$  & {\cellcolor{blue!0}} $-4.896 \times 10^{-4}$ & {\cellcolor{blue!0}} $-3.215 \times 10^{-5}$ & {\cellcolor{blue!25}} -0.03007 & {\cellcolor{blue!1}} -0.002045 & {\cellcolor{red!5}} 0.005920 & {\cellcolor{red!40}} 0.04709 \\
$\gamma$  & {\cellcolor{red!0}} $1.467 \times 10^{-4}$ & {\cellcolor{red!11}} 0.004489 & {\cellcolor{blue!40}} -0.01515 & {\cellcolor{red!0}} $1.824 \times 10^{-4}$ & {\cellcolor{red!10}} 0.004164 & {\cellcolor{red!21}} 0.008094 \\
$f_{\text{ice}}$  & {\cellcolor{blue!0}} $-3.694 \times 10^{-5}$ & {\cellcolor{blue!37}} -0.005475 & {\cellcolor{blue!37}} -0.005348 & {\cellcolor{red!2}} $3.690 \times 10^{-4}$ & {\cellcolor{blue!16}} -0.002356 & {\cellcolor{red!40}} 0.005773 \\
$r_{\text{effw}}$ [$\mu m$] & {\cellcolor{red!0}} 0.001788 & {\cellcolor{blue!9}} -0.05233 & {\cellcolor{red!29}} 0.1663 & {\cellcolor{blue!0}} -0.001699 & {\cellcolor{blue!4}} -0.02809 & {\cellcolor{red!40}} 0.2275 \\
$r_{\text{effi}}$ [$\mu m$] & {\cellcolor{red!1}} 0.02186 & {\cellcolor{blue!21}} -0.3809 & {\cellcolor{blue!21}} -0.3765 & {\cellcolor{red!0}} 0.01551 & {\cellcolor{blue!23}} -0.4141 & {\cellcolor{blue!40}} -0.7065 \\
\bottomrule
\end{tabular}

\end{adjustbox}
\caption{Energy Score (ES), Root Mean Squared Error (RMSE), Pearson R coefficient between the output variables' means and the corresponding ground truth and mean bias for the relevant model versions. ES is shown for the multivariate models. See Table~\ref{tab:models} for a summary of the model features. The cell shading is row-wise normalized, and proportional to the cell value. In the bias section, red indicates positive bias values, white zero values and blue negative values.}
\label{tab:global_metrics}
\end{table}

\subsection{Distributions of Predicted Means}

\begin{figure}
    \centering
    \includegraphics[width=0.75\linewidth]{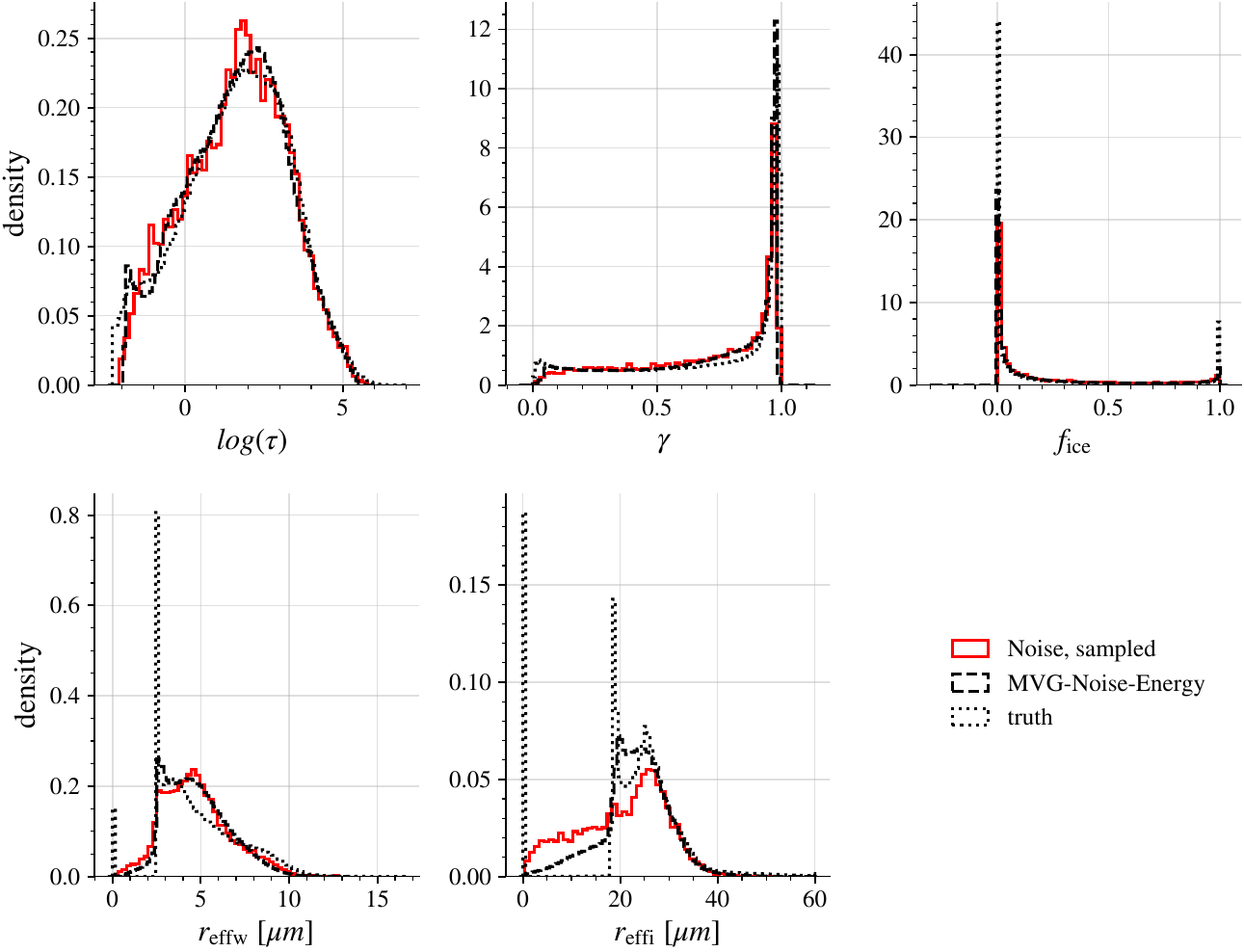}
    \caption{Histograms for the NN output variables, comparing the means of the sampled conditional distributions assuming noisy inputs (red), the predicted means from the MVG-Noise-Energy model (black, dashed) and the ground truth (black, dotted). In the panels for water and ice effective radii's means, ice fraction is used to filter out cases with ice-only ($f_{\mathrm{ice}}\geq 0.9$) and water-only ($f_{\mathrm{ice}}<0.01$), respectively. The filtering is based on the predicted mean ice fraction for MVG-Noise-Energy and MVG-Energy models, and on the sampled mean of ice fraction for the sampled model.}
    \label{fig:means_hist}
\end{figure}

In this section, we assess the predicted means from the MVG-Noise-Energy model. In this and in the following Sections~\ref{sec:pred_cov}, \ref{sec:gaussianity} and~\ref{sec:uncertainty_stratification}, a filtering based on ice fraction is applied to water and ice effective radii, in order to exclude the effective radii associated to a cloud water phase that is absent. For this purpose, only-liquid-water cases are defined by $f_{\mathrm{ice}}<0.01$ and only-ice cases by $f_{\mathrm{ice}}\geq 0.9$, matching the thresholds employed in ~\ref{sec:situation_analysis}. For the BO predictions, the filtering is based on the BO ice fraction predicted mean, whereas, for the sampled predictive distribution, the filtering is based on the ice fraction sampled mean. The distributions of predicted means closely match the target truths and the sampled means (Fig.~\ref{fig:means_hist}), with the sampled and predicted means being closer together, especially in areas where the distributions show sharp features or boundaries. Such regions correspond to areas where the underlying conditional distribution is expected to be skewed and less Gaussian (see Section~\ref{sec:gaussianity} below), and where the predicted mean on average is not expected to be a good proxy for the truth. These findings are also supported by the Pearson R reported in Table~\ref{tab:PearsonR_with_sampled}, which shows a higher correlation between the predicted and sampled mean. The global mean bias already shown in Table~\ref{tab:global_metrics} is very close to zero for all variables. These results show that the first moment was correctly and reliably learned during the training.

\begin{table}
\centering
\begin{adjustbox}{max width=\textwidth}
    \begin{tabular}{lrrr}
\toprule
 & R$\left[x, \mu\right]$ & R$\left[\mu_{\text{s}}, \mu\right]$ & R$\left[\mu_{\text{s}}, x\right]$ \\
\midrule
$\log(\tau)$  & {\cellcolor{lightblue!15}} 0.949 & {\cellcolor{lightblue!100}} 0.977 & {\cellcolor{lightblue!57}} 0.963 \\
$\gamma$  & {\cellcolor{lightblue!15}} 0.878 & {\cellcolor{lightblue!100}} 0.932 & {\cellcolor{lightblue!57}} 0.905 \\
$f_{\text{ice}}$  & {\cellcolor{lightblue!15}} 0.867 & {\cellcolor{lightblue!37}} 0.880 & {\cellcolor{lightblue!100}} 0.916 \\
$r_{\text{effw}}$ [$\mu m$] & {\cellcolor{lightblue!15}} 0.810 & {\cellcolor{lightblue!100}} 0.901 & {\cellcolor{lightblue!78}} 0.878 \\
$r_{\text{effi}}$ [$\mu m$] & {\cellcolor{lightblue!15}} 0.781 & {\cellcolor{lightblue!100}} 0.864 & {\cellcolor{lightblue!34}} 0.802 \\
\bottomrule
\end{tabular}
\end{adjustbox}
\vspace{2mm}
\caption{Pearson R for the MVG-Noise-Energy model, computed on the pairs $(x, \mu)$, $(\mu_{\mathrm{s}}, \mu)$, $(\mu_{\mathrm{s}}, x)$, where $x$ denotes the ground truth for a given output variable, $\mu$ the corresponding predicted mean, and $\mu_{\mathrm{s}}$ the sampled mean. The sampled mean is computed with the technique described in Section~\ref{section:sampling_covariances}. The cell shading is row-wise normalized, and proportional to the cell value.}
\label{tab:PearsonR_with_sampled}
\end{table}

\subsection{Distributions of Predicted Covariances}\label{sec:pred_cov}

\begin{figure}
    \centering
    \includegraphics[width=0.65\linewidth]{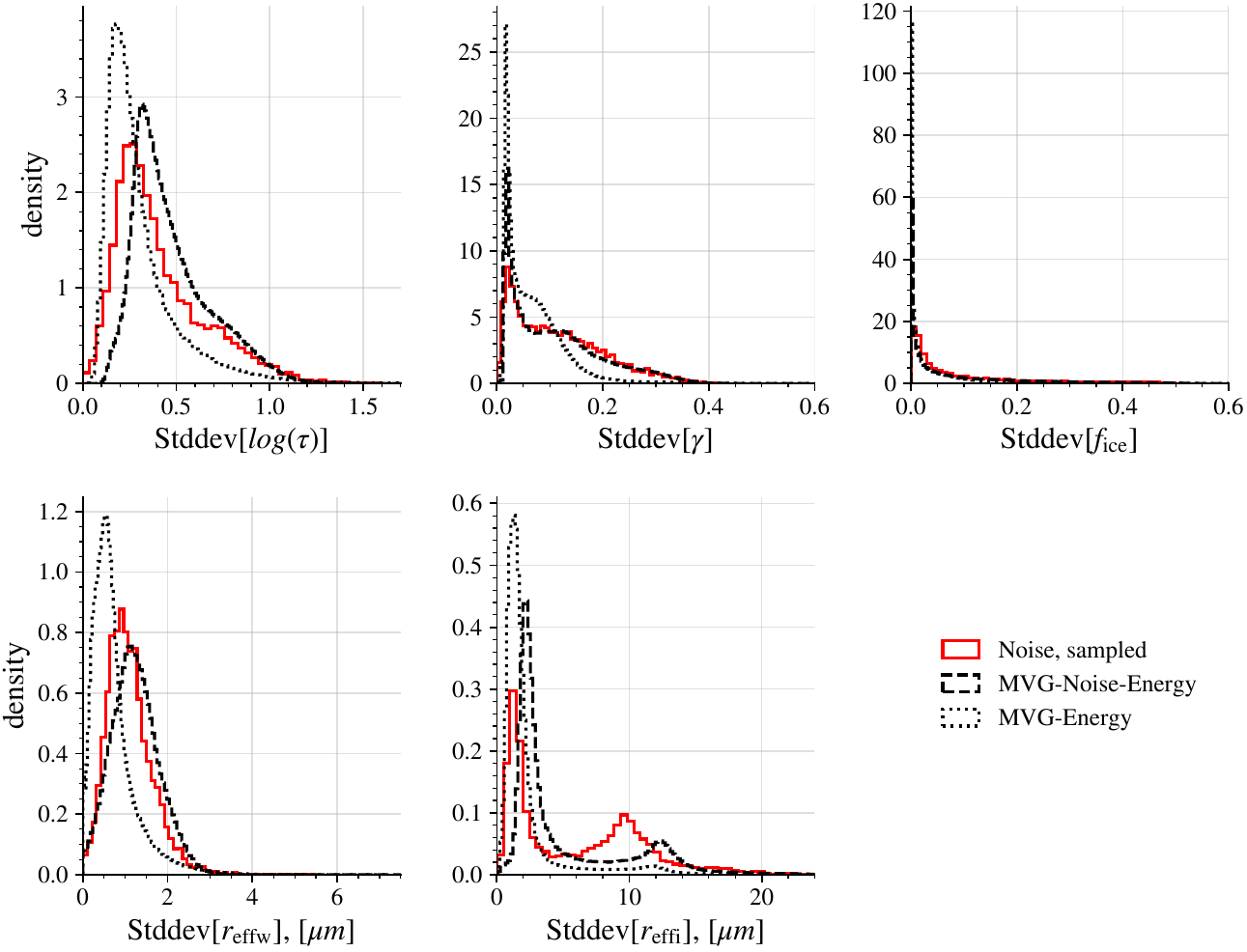}
    \caption{Histograms for the NN output variables, comparing the standard deviations of the sampled conditional distributions assuming noisy inputs (red), the predicted standard deviations from the MVG-Noise-Energy model (black, dashed) and the predicted standard deviations from the MVG-Energy model (black, dotted). In the panels for water and ice effective radii's uncertainties, ice fraction is used to filter out cases with ice-only ($f_{\mathrm{ice}}\geq 0.9$) and water-only ($f_{\mathrm{ice}}<0.01$), respectively. The filtering is based on the predicted mean ice fraction for MVG-Noise-Energy and MVG-Energy models, and on the sampled mean of ice fraction for the sampled model.}
    \label{fig:sigmas_hist}
\end{figure}

\begin{figure}
    \centering
    \includegraphics[width=0.65\linewidth]{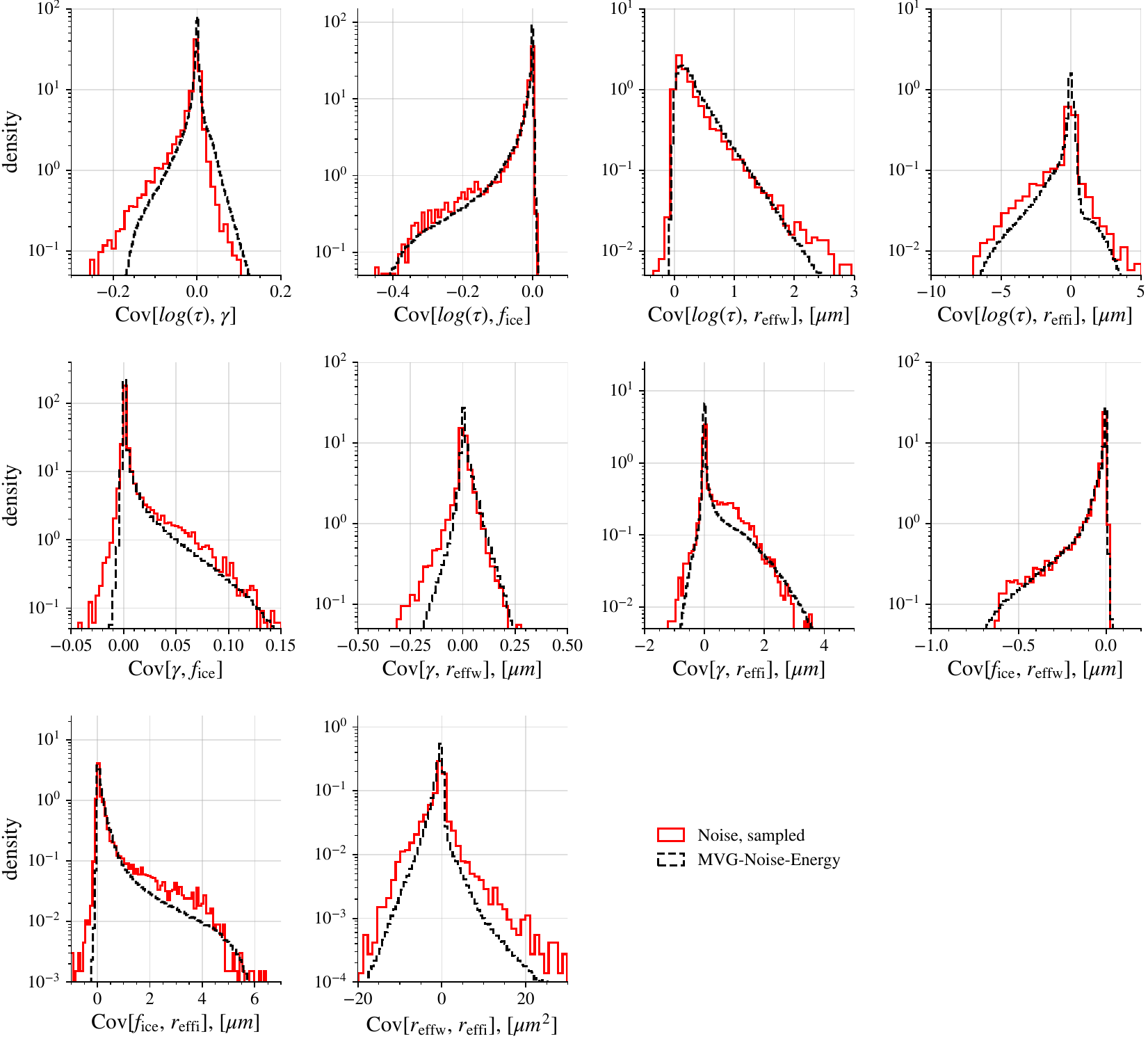}
    \caption{Histograms for the NN output variables, comparing the off-diagonal covariances of the sampled conditional distributions assuming noisy inputs (red) and the off-diagonal predicted covariances from the MVG-Noise-Energy model (black, dashed). In the panels including water and/or ice effective radii, ice fraction is used to filter out cases with ice-only ($f_{\mathrm{ice}}\geq 0.9$) and water-only ($f_{\mathrm{ice}}<0.01$), respectively. The filtering is based on the predicted mean ice fraction for MVG-Noise-Energy model, and on the sampled mean of ice fraction for the sampled model.}
    \label{fig:cov_hist}
\end{figure}

In this section, the covariance structure is presented. Let us start by considering uncertainties in terms of standard deviations, corresponding to the square root of the diagonal part of the covariance matrix. In Fig.~\ref{fig:sigmas_hist}, the distributions of predicted and sampled uncertainties from the ES-based models are compared. In general, a good agreement is found, the only exception being the uncertainties of ice effective radius, for which the location of the peaks of the distribution is slightly underestimated and their size not as accurately represented.  Nevertheless, the range of values and the bimodal character of distribution are correctly reproduced. As a reference, on the same plot the histogram of the predicted uncertainties for the MVG-Energy model are superimposed. The comparison between MVG-Energy and MVG-Noise-Energy highlights the effect of the noise injection on the predicted uncertainties, which tends to produce higher uncertainties. At the same time, it clarifies that input noise is not the main source of output uncertainty, as the version without noise tends to produce lower uncertainties, yet on the same order of magnitude. Looking then at the off-diagonal elements of covariance in Fig.~\ref{fig:cov_hist}, the sampled and predicted covariance histograms have a good overlap. In particular, the predictions agree on the sign of covariances, especially for the pairs of variables expected to have a definite covariance sign (either positive or negative) from the sampled covariance histogram. Overall, the findings above indicate a good consistency between the learned and the sampled covariance structure. A more comprehensive analysis of predictions in a probabilistic sense follows in the next section.

\subsection{Testing Gaussianity and Uncertainty}\label{sec:gaussianity}

\begin{figure}
    \centering
    \includegraphics[width=0.75\linewidth]{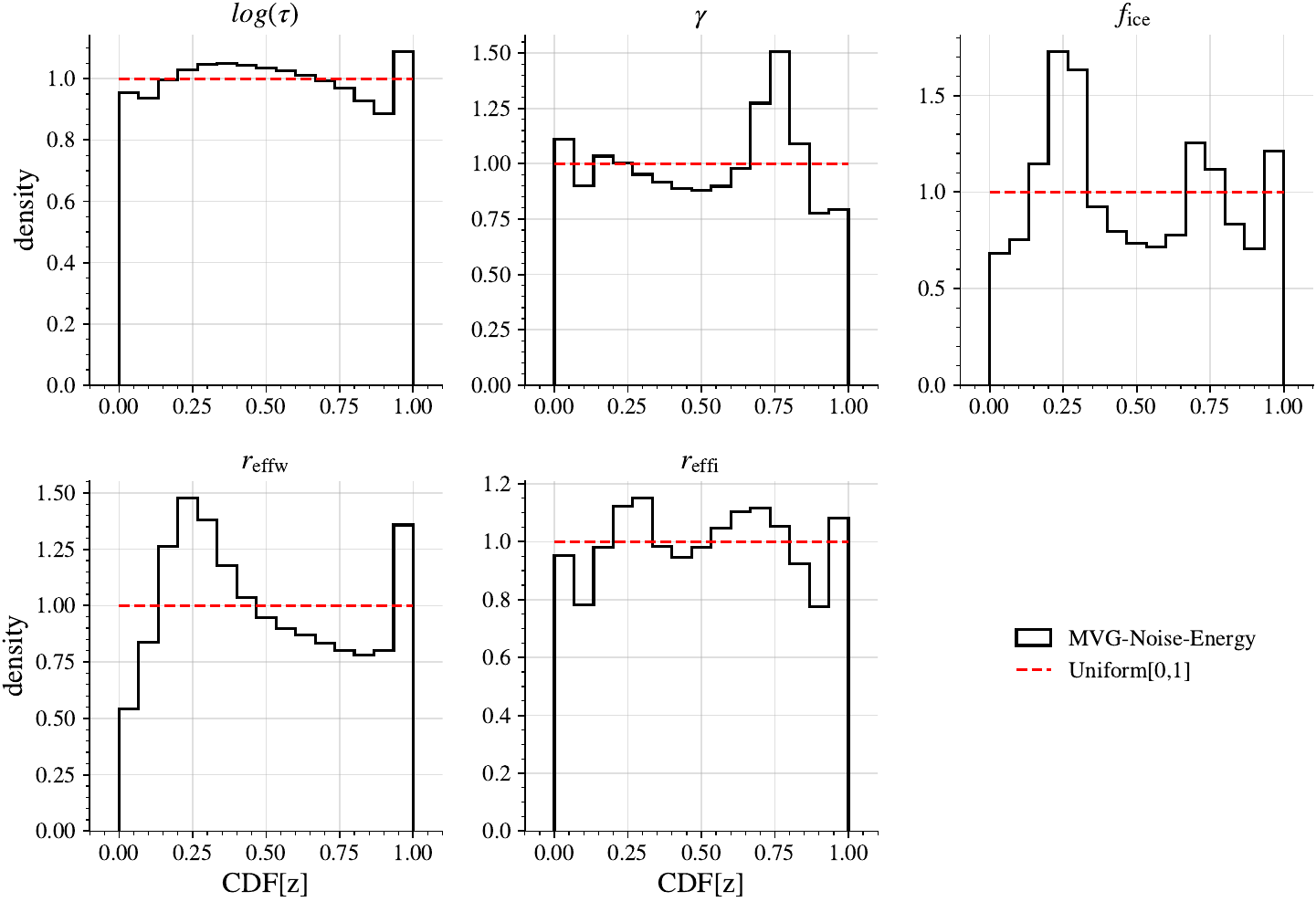}
    \caption{PIT plot for the MVG-Noise-Energy model (black). As a reference, the uniform distribution $U[0,1]$ between 0 and 1 is shown (red, dashed). In the panels for water and ice effective radii, ice fraction is used to filter out cases with ice-only ($f_{\mathrm{ice}}\geq 0.9$) and water-only ($f_{\mathrm{ice}}<0.01$), respectively, based on the predicted mean.}
    \label{fig:PIT}
\end{figure}

\begin{figure}
    \centering
    \includegraphics[width=0.75\linewidth]{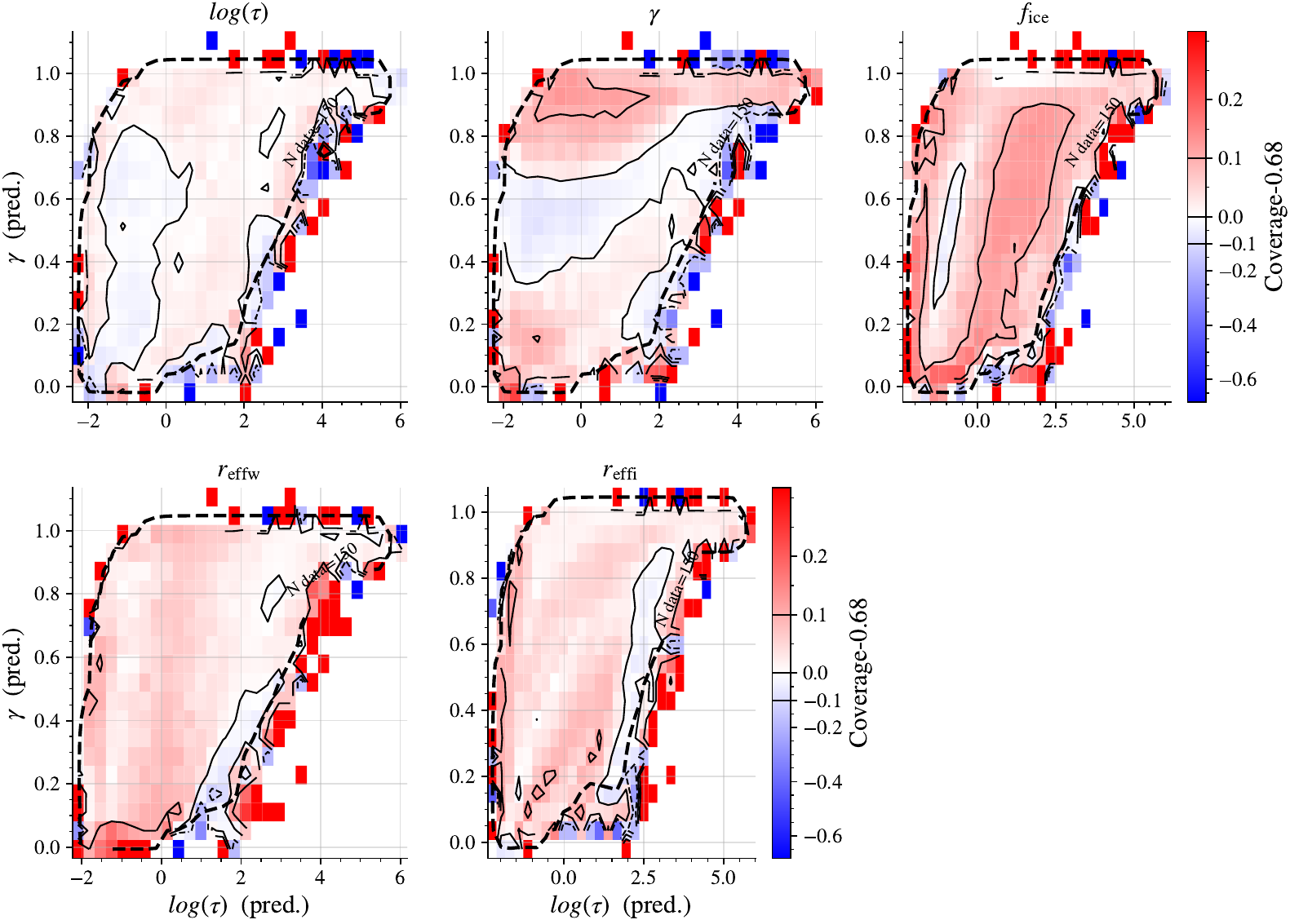}
    \caption{Departures from the nominal value of the 1--$\sigma$ coverage of the NN output variables, conditioned to the predicted optical thickness $\log(\tau)$ and column cloud fraction $\gamma$ for the MVG-Noise-Energy model. The thick, dashed line is defined by the condition $N_{bins}=150$, $N_{bins}$ being the number of test points in the given bin.  To compute water and ice effective radii's coverages, ice fraction is used to filter out cases with ice-only ($f_{\mathrm{ice}}\geq 0.9$) and water-only ($f_{\mathrm{ice}}<0.01$), respectively, based on the predicted mean.}
    \label{fig:coveragestrat_MVGNoiseEnergy}
\end{figure}

Using the methods introduced in Section~\ref{sec:model_performance}, we assess the calibration of predictions and the quality of the Gaussianity hypothesis, focusing on the univariate Gaussian distributions resulting from the marginals of the full multivariate output distribution. We first show the PIT plots in Fig.~\ref{fig:PIT}. The histograms show that $\log(\tau)$ is the closest to Gaussianity, which a-posteriori justifies choosing the logarithm of $\tau$ as the output variable. As a consequence, $\tau$ is well described by the associated log-normal distribution. If needed, the mean and variance of the latter can be easily computed from the predicted mean and standard deviations using standard formulae. Considering the other output variables, boundedness and sharp peaks in their distributions can be indicated as the main reasons for the deviation from Gaussianity.
To assess the reliability of the uncertainty estimates, global 1--$\sigma$ coverages for all variables are reported in Table~\ref{tab:coverage_global}. Overall, coverage is well-behaved, being reasonably close to the nominal value of 0.68. It can be noticed, however, that the model has a tendency to over-cover, i.e., given the negligible bias shown above, to slightly over-estimate the output uncertainties. In accordance with the PIT plot, $\log(\tau)$ is the closest to the nominal value. The deviation of the 1--$\sigma$ coverage from the nominal value is then shown in Fig.~\ref{fig:coveragestrat_MVGNoiseEnergy}, conditioned to mean predictions in the $(\gamma,\tau)$ plane. On the same plot, the region with bins containing more than 150 test points is marked (thick dashed line). The bins outside the area clearly show higher positive and negative departures from the nominal coverage, suggesting, in case of statistically significant departures, that the lower amount of points in the tails of the training distributions could not constrain the model predictions reliably during the training phase. As for the inner region, consistently with the global behaviour, a predominantly weak positive departure is found. Overall, these findings show a good calibration of the probabilistic predictions, with reliable uncertainty estimates at 1--$\sigma$.

\begin{table}
\centering
\begin{adjustbox}{max width=\textwidth}
    \begin{tabular}{lr}
\toprule
 & 1--$\sigma$ Coverage \\
\midrule
$\log(\tau)$  & 0.6940 \\
$\gamma$  & 0.7207 \\
$f_{\text{ice}}$  & 0.7609 \\
$r_{\text{effw}}$ [$\mu m$] & 0.7158 \\
$r_{\text{effi}}$ [$\mu m$] & 0.7151 \\
\bottomrule
\end{tabular}
\end{adjustbox}
\caption{Global 1--$\sigma$ coverage of the NN output variables for the MVG-Noise-Energy model. To compute water and ice effective radii's coverages, ice fraction is used to filter out cases with ice-only ($f_{\mathrm{ice}}\geq 0.9$) and water-only ($f_{\mathrm{ice}}<0.01$), respectively, based on the predicted mean.}
\label{tab:coverage_global}
\end{table}

\subsection{Uncertainty Stratification}\label{sec:uncertainty_stratification}

\begin{figure}
    \centering
    \includegraphics[width=0.95\linewidth]{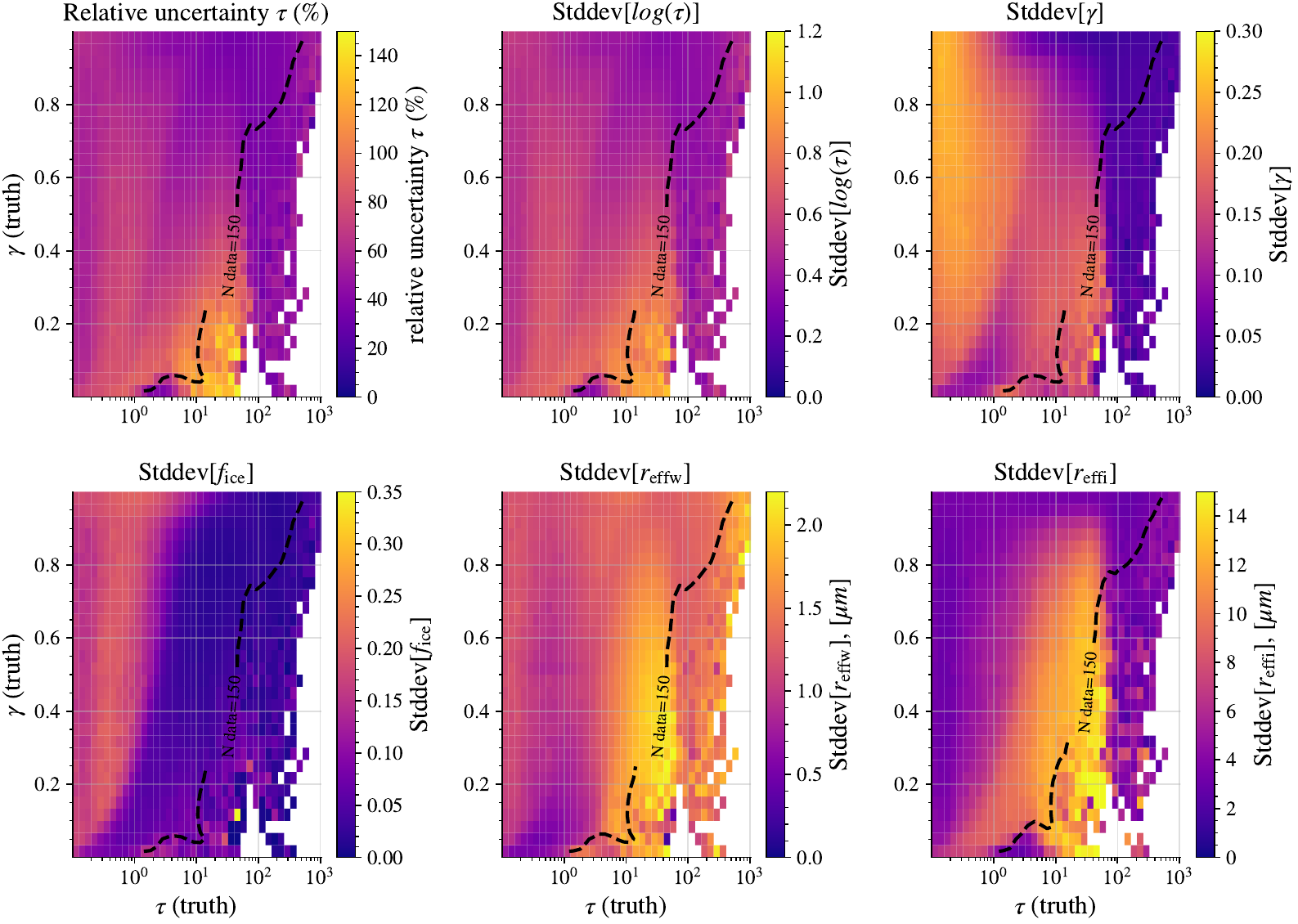}
    \caption{Output uncertainties conditioned to the ground truth total optical thickness $\tau$ and column cloud fraction $\gamma$ for the MVG-Noise-Energy model. The thick, dashed line is defined by the condition $N_{bins}=150$, $N_{bins}$ being the number of test points in the given bin. In the panels for water and ice effective radii's uncertainties, ice fraction is used to filter out cases with ice-only ($f_{\mathrm{ice}}\geq 0.9$) and water-only ($f_{\mathrm{ice}}<0.01$), respectively, based on the predicted mean.}
    \label{fig:sigmastrat_taugamma_MVGNoNoiseEnergy}
\end{figure}

In Fig.~\ref{fig:sigmastrat_taugamma_MVGNoNoiseEnergy} we present a stratification of the predicted uncertainties, focusing for simplicity on its dependence on the total optical thickness $\tau$ and the column cloud fraction $\gamma$, as these variables provide a good description of the cloud scene. In this case, the test data are conditioned with respect to the ground truth, as here we focus on how the invertibility behaves given the actual physical situation. Similarly to Fig.~\ref{fig:coveragestrat_MVGNoiseEnergy}, the region with bins containing more than 150 test points is highlighted as a reference. For convenience, the relative uncertainty $\tau/\text{Stddev}[\tau]$ is shown as a percentage computed from the mean and square root of the variance of the associated log-normal distribution. As can be verified with a Taylor expansion, the relative uncertainty is well approximated by $\text{Stddev}[log(\tau)]$, as the latter exhibits relatively small values. Despite its more intuitive physical interpretation, it must be stressed that $\text{Stddev}[\tau]$ does not have a constant coverage, as it is distributed log-normally. The figure shows that $\tau$ relative uncertainty is lower in thicker clouds and for higher column cloud fractions $\gamma$. In this scenario, the relative uncertainty reaches values as low as $\approx 40\%$. The same pattern can be found in $\gamma$. A physical reason for this behaviour could be that for low values of $\tau$ or $\gamma$, the cloud signal is weak, compared to the surface signal. When $\tau$ or $\gamma$ are increased, the signal to noise improves and better estimates are possible. $f_{\mathrm{ice}}$ is characterized by a wider, low uncertainty region, extending from low ($\tau$, $\gamma$) values to the thick and fully cloudy scenes. The effective radii show a behaviour that is remarkably different from the other variables: there is a clear tendency for uncertainty to increase with optical depth. This tendency is probably related to the fact that much of the information on the effective radii is contained in the angular distribution of the scattered photons, which is different for each of the six channels. For thin clouds, in which single scattering dominates, the BO can use the signal in the different channels related to the radii efficiently. For thicker clouds, multiple scattering dominates the reflectances. As the multiple scattering contribution to reflectance contains almost no information on the angular distribution\footnote{The photon direction is known only for the first scattering process.}, the signal to noise ratio for effective radii is reduced for higher optical depths, increasing the uncertainty. However, additional information on the radii is available from the near-infrared channels, where even in the multiple scattering dominated regime, absorption in the cloud particles causes reflectance to depend on the effective radii. The differences in the two NIR channels are stronger for larger radii, which are typically found for higher optical depths. This second source of information on the radii could be responsible for the relatively low uncertainties of $r_{\mathbf{effi}}$ at high optical depths.

\subsubsection{Situation Analysis}\label{sec:situation_analysis}

In this section, focusing on MVG-Noise-Energy model, we analyse the predicted uncertainties and their calibration using a physically based set of situations. The classification is based on the ground truth column cloud fraction $\gamma$, total optical thickness $\tau$ and ice fraction $f_{\mathrm{ice}}$. The set of situations is defined from the Cartesian product of the cases defined for each variable. As for $\gamma$, we distinguish between fully cloudy ($\gamma\geq0.95$) and partially cloudy ($\gamma<0.95$). Three cases are then defined in terms of $\tau$, namely, thick clouds ("thick", $log(\tau)\geq 2.3$), intermediate clouds ("inter", $1.1\leq log(\tau)<2.3$), and thin clouds ("thin", $log(\tau)<1.1$). Finally, the water phase content is classified into three categories, namely, liquid water clouds ("water", $f_{\mathrm{ice}}<0.01$), clouds containing both liquid and solid water ("w+i", $0.01\leq f_{\mathrm{ice}}<0.9$), and ice clouds ("ice", $f_{\mathrm{ice}}\geq 0.9$). In the first row of Table~\ref{tab:situations_sigma}, the fractions of points in the test dataset for each situation are shown. In the rest of Table~\ref{tab:situations_sigma}, the mean predicted uncertainties for all output variables are presented (the bin mean is used). The uncertainty associated with the total optical thickness $\tau$ confirms the behaviour outlined in Section~\ref{sec:uncertainty_stratification}, being a decreasing function of total optical thickness, and with sharper predictions in presence of fully cloudy cases. Here, in addition, $\tau$'s relative uncertainty is shown to decrease as a function of ice fraction, the only exception being the fully cloudy, thick cloud situations. While sharing with $\tau$ the same dependence on cloudiness and thickness, the behaviour of $\gamma$'s uncertainty as a function of $f_{\mathrm{ice}}$ is different, and not as consistent: overall, it tends to be non-monotonically decreasing, with the exception of the partially cloudy, thin cases, exhibiting a monotonically increasing behaviour. Considering ice fraction's uncertainty, it is mostly sensitive to ice fraction itself, with lowest values for only-water cases, and saturating to uncertainties $\approx0.1\div0.2$ in the w+i and only-ice cases. A weak residual dependence on thickness can be found, again with a tendency for sharper predictions towards thicker clouds. The uncertainty in water effective radius shows more complex behaviour, particularly as a function of ice fraction, with the lowest values found for thin, partially cloudy cases, consistent with Figure~\ref{fig:sigmastrat_taugamma_MVGNoNoiseEnergy}. Finally, the uncertainty in ice effective radius decreases with increasing ice fraction. Table~\ref{tab:situations_coverage} reports the $1$-$\sigma$ coverage for the situation-based uncertainty stratification presented in this section. While overall calibration is reasonably good, a few situations show significant departures from the nominal value of $0.68$, which can be grouped into three categories. The first group includes very infrequent situations (with fewer than $0.4\%$ of points), suggesting that, where departures are statistically significant, the limited number of points in the tails of the training distributions may not have been sufficient to reliably constrain the uncertainty predictions during training. The second group, characterised by over-coverage in ice fraction or column cloud fraction, corresponds to more frequent situations that sample the boundary of their bounded distribution. As discussed in Section~\ref{sec:gaussianity}, the Gaussianity assumption is not ideal in the presence of sharp features and bounded variables. Finally, the third group contains only the "partially cloudy, thick, w+i" situation (representing $5.21\%$ of the test dataset), in which column cloud fraction is found to be significantly under-covering ($1$-$\sigma$ coverage of $0.443$), meaning that uncertainties on column cloud fraction tend to be underestimated for this situation.

\begin{table}
\centering
\begin{adjustbox}{max width=\textwidth}
    \begin{tabular}{c|
    ccc|ccc|ccc|
    ccc|ccc|ccc
}
\toprule
\rowcolor{white}
cloudiness 
& \multicolumn{9}{c|}{fully cloudy}
& \multicolumn{9}{c}{partially cloudy} \\
\cmidrule(lr){2-10}
\cmidrule(lr){11-19}

\rowcolor{white}
thickness 
& \multicolumn{3}{c}{thick}
& \multicolumn{3}{c}{inter}
& \multicolumn{3}{c|}{thin}
& \multicolumn{3}{c}{thick}
& \multicolumn{3}{c}{inter}
& \multicolumn{3}{c}{thin} \\
\cmidrule(lr){1-1}
\cmidrule(lr){2-4}
\cmidrule(lr){5-7}
\cmidrule(lr){8-10}
\cmidrule(lr){11-13}
\cmidrule(lr){14-16}
\cmidrule(lr){17-19}

\rowcolor{white}
phase 
& water & w+i & ice
& water & w+i & ice
& water & w+i & ice
& water & w+i & ice
& water & w+i & ice
& water & w+i & ice \\
\midrule
Percentage [\%] & {\cellcolor{lightblue!43}} 4.95 & {\cellcolor{lightblue!100}} 15.0 & {\cellcolor{lightblue!15}} 0.0201 & {\cellcolor{lightblue!15}} 0.104 & {\cellcolor{lightblue!57}} 7.47 & {\cellcolor{lightblue!17}} 0.519 & {\cellcolor{lightblue!15}} 0.00547 & {\cellcolor{lightblue!42}} 4.90 & {\cellcolor{lightblue!31}} 2.92 & {\cellcolor{lightblue!83}} 12.1 & {\cellcolor{lightblue!44}} 5.21 & {\cellcolor{lightblue!15}} 0.00490 & {\cellcolor{lightblue!88}} 12.9 & {\cellcolor{lightblue!42}} 4.79 & {\cellcolor{lightblue!16}} 0.338 & {\cellcolor{lightblue!92}} 13.7 & {\cellcolor{lightblue!61}} 8.25 & {\cellcolor{lightblue!54}} 6.89 \\
\midrule
rel. err. $\tau\,\,[\%]$  & {\cellcolor{lightblue!34}} 34.8 & {\cellcolor{lightblue!39}} 37.9 & {\cellcolor{lightblue!25}} 29.0 & {\cellcolor{lightblue!49}} 44.7 & {\cellcolor{lightblue!49}} 44.1 & {\cellcolor{lightblue!18}} 24.2 & {\cellcolor{lightblue!69}} 57.0 & {\cellcolor{lightblue!66}} 55.0 & {\cellcolor{lightblue!48}} 43.9 & {\cellcolor{lightblue!52}} 46.1 & {\cellcolor{lightblue!47}} 42.9 & {\cellcolor{lightblue!19}} 25.1 & {\cellcolor{lightblue!71}} 58.2 & {\cellcolor{lightblue!63}} 53.3 & {\cellcolor{lightblue!15}} 22.1 & {\cellcolor{lightblue!100}} 76.9 & {\cellcolor{lightblue!92}} 72.4 & {\cellcolor{lightblue!56}} 49.0 \\
$\log(\tau)$  & {\cellcolor{lightblue!38}} 0.336 & {\cellcolor{lightblue!43}} 0.363 & {\cellcolor{lightblue!27}} 0.280 & {\cellcolor{lightblue!54}} 0.419 & {\cellcolor{lightblue!52}} 0.411 & {\cellcolor{lightblue!18}} 0.234 & {\cellcolor{lightblue!73}} 0.519 & {\cellcolor{lightblue!67}} 0.492 & {\cellcolor{lightblue!50}} 0.399 & {\cellcolor{lightblue!56}} 0.430 & {\cellcolor{lightblue!50}} 0.402 & {\cellcolor{lightblue!20}} 0.244 & {\cellcolor{lightblue!73}} 0.523 & {\cellcolor{lightblue!66}} 0.482 & {\cellcolor{lightblue!15}} 0.215 & {\cellcolor{lightblue!100}} 0.660 & {\cellcolor{lightblue!93}} 0.624 & {\cellcolor{lightblue!58}} 0.442 \\
$\gamma$ & {\cellcolor{lightblue!22}} 0.0476 & {\cellcolor{lightblue!15}} 0.0312 & {\cellcolor{lightblue!17}} 0.0362 & {\cellcolor{lightblue!57}} 0.120 & {\cellcolor{lightblue!27}} 0.0579 & {\cellcolor{lightblue!28}} 0.0590 & {\cellcolor{lightblue!82}} 0.171 & {\cellcolor{lightblue!61}} 0.128 & {\cellcolor{lightblue!77}} 0.161 & {\cellcolor{lightblue!50}} 0.104 & {\cellcolor{lightblue!33}} 0.0696 & {\cellcolor{lightblue!26}} 0.0556 & {\cellcolor{lightblue!72}} 0.150 & {\cellcolor{lightblue!62}} 0.129 & {\cellcolor{lightblue!52}} 0.109 & {\cellcolor{lightblue!55}} 0.115 & {\cellcolor{lightblue!92}} 0.191 & {\cellcolor{lightblue!100}} 0.207 \\
$f_{\text{ice}}$ & {\cellcolor{lightblue!15}} 0.00931 & {\cellcolor{lightblue!38}} 0.0659 & {\cellcolor{lightblue!88}} 0.187 & {\cellcolor{lightblue!20}} 0.0216 & {\cellcolor{lightblue!69}} 0.141 & {\cellcolor{lightblue!77}} 0.160 & {\cellcolor{lightblue!26}} 0.0365 & {\cellcolor{lightblue!100}} 0.215 & {\cellcolor{lightblue!80}} 0.167 & {\cellcolor{lightblue!17}} 0.0149 & {\cellcolor{lightblue!30}} 0.0479 & {\cellcolor{lightblue!79}} 0.164 & {\cellcolor{lightblue!23}} 0.0293 & {\cellcolor{lightblue!45}} 0.0823 & {\cellcolor{lightblue!65}} 0.132 & {\cellcolor{lightblue!32}} 0.0527 & {\cellcolor{lightblue!91}} 0.193 & {\cellcolor{lightblue!70}} 0.144 \\
$r_{\text{effw}}$ [$\mu m$] & {\cellcolor{lightblue!68}} 1.23 & {\cellcolor{lightblue!83}} 1.35 & - & {\cellcolor{lightblue!86}} 1.38 & {\cellcolor{lightblue!84}} 1.37 & - & {\cellcolor{lightblue!79}} 1.32 & {\cellcolor{lightblue!62}} 1.18 & - & {\cellcolor{lightblue!100}} 1.50 & {\cellcolor{lightblue!87}} 1.39 & - & {\cellcolor{lightblue!82}} 1.35 & {\cellcolor{lightblue!81}} 1.34 & - & {\cellcolor{lightblue!15}} 0.769 & {\cellcolor{lightblue!42}} 1.00 & - \\
$r_{\text{effi}}$ [$\mu m$] & - & {\cellcolor{lightblue!33}} 3.57 & {\cellcolor{lightblue!19}} 2.57 & - & {\cellcolor{lightblue!36}} 3.80 & {\cellcolor{lightblue!15}} 2.33 & - & {\cellcolor{lightblue!41}} 4.13 & {\cellcolor{lightblue!25}} 3.01 & - & {\cellcolor{lightblue!85}} 7.23 & {\cellcolor{lightblue!17}} 2.46 & - & {\cellcolor{lightblue!85}} 7.23 & {\cellcolor{lightblue!15}} 2.28 & - & {\cellcolor{lightblue!62}} 5.62 & {\cellcolor{lightblue!25}} 3.02 \\
\bottomrule
\end{tabular}

\end{adjustbox}
\caption{Predicted uncertainties for the set of defined situations (see Section~\ref{sec:situation_analysis} for the definitions). The mean within bin is shown. The fraction of points [\%] in the test dataset in each situation is reported in the first row. The cell shading is row-wise normalized, and proportional to the cell value. For ice and water effective radii, "water" and "ice" situations are not shown, respectively, as the associated cloud water phase is absent.}
\label{tab:situations_sigma}
\end{table}

\begin{table}
\centering
\begin{adjustbox}{max width=\textwidth}
    \begin{tabular}{c|
    ccc|ccc|ccc|
    ccc|ccc|ccc
}
\toprule
\rowcolor{white}
cloudiness 
& \multicolumn{9}{c|}{fully cloudy}
& \multicolumn{9}{c}{partially cloudy} \\
\cmidrule(lr){2-10}
\cmidrule(lr){11-19}

\rowcolor{white}
thickness 
& \multicolumn{3}{c}{thick}
& \multicolumn{3}{c}{inter}
& \multicolumn{3}{c|}{thin}
& \multicolumn{3}{c}{thick}
& \multicolumn{3}{c}{inter}
& \multicolumn{3}{c}{thin} \\
\cmidrule(lr){1-1}
\cmidrule(lr){2-4}
\cmidrule(lr){5-7}
\cmidrule(lr){8-10}
\cmidrule(lr){11-13}
\cmidrule(lr){14-16}
\cmidrule(lr){17-19}

\rowcolor{white}
phase 
& water & w+i & ice
& water & w+i & ice
& water & w+i & ice
& water & w+i & ice
& water & w+i & ice
& water & w+i & ice \\
\midrule
Percentage [\%] & {\cellcolor{lightblue!43}} 4.95 & {\cellcolor{lightblue!100}} 15.0 & {\cellcolor{lightblue!15}} 0.0201 & {\cellcolor{lightblue!15}} 0.104 & {\cellcolor{lightblue!57}} 7.47 & {\cellcolor{lightblue!17}} 0.519 & {\cellcolor{lightblue!15}} 0.00547 & {\cellcolor{lightblue!42}} 4.90 & {\cellcolor{lightblue!31}} 2.92 & {\cellcolor{lightblue!83}} 12.1 & {\cellcolor{lightblue!44}} 5.21 & {\cellcolor{lightblue!15}} 0.00490 & {\cellcolor{lightblue!88}} 12.9 & {\cellcolor{lightblue!42}} 4.79 & {\cellcolor{lightblue!16}} 0.338 & {\cellcolor{lightblue!92}} 13.7 & {\cellcolor{lightblue!61}} 8.25 & {\cellcolor{lightblue!54}} 6.89 \\
\midrule
$\tau$  & {\cellcolor{lightgreen}} 0.711 & {\cellcolor{lightgreen}} 0.732 & {\cellcolor{lightgreen}} 0.775 & {\cellcolor{lightgreen}} 0.627 & {\cellcolor{yellow}} 0.788 & {\cellcolor{yellow}} 0.785 & {\cellcolor{salmon}} 0.405 & {\cellcolor{yellow}} 0.800 & {\cellcolor{yellow}} 0.809 & {\cellcolor{lightgreen}} 0.699 & {\cellcolor{lightgreen}} 0.684 & {\cellcolor{lightgreen}} 0.710 & {\cellcolor{yellow}} 0.803 & {\cellcolor{yellow}} 0.799 & {\cellcolor{lightgreen}} 0.764 & {\cellcolor{yellow}} 0.850 & {\cellcolor{yellow}} 0.812 & {\cellcolor{yellow}} 0.795 \\
$\log(\tau)$  & {\cellcolor{lightgreen}} 0.679 & {\cellcolor{lightgreen}} 0.694 & {\cellcolor{lightgreen}} 0.710 & {\cellcolor{yellow}} 0.555 & {\cellcolor{lightgreen}} 0.725 & {\cellcolor{lightgreen}} 0.756 & {\cellcolor{salmon}} 0.312 & {\cellcolor{lightgreen}} 0.683 & {\cellcolor{lightgreen}} 0.728 & {\cellcolor{lightgreen}} 0.651 & {\cellcolor{lightgreen}} 0.639 & {\cellcolor{lightgreen}} 0.684 & {\cellcolor{lightgreen}} 0.732 & {\cellcolor{lightgreen}} 0.730 & {\cellcolor{lightgreen}} 0.743 & {\cellcolor{lightgreen}} 0.715 & {\cellcolor{lightgreen}} 0.665 & {\cellcolor{lightgreen}} 0.676 \\
$\gamma$ & {\cellcolor{yellow}} 0.814 & {\cellcolor{yellow}} 0.829 & {\cellcolor{salmon}} 0.948 & {\cellcolor{salmon}} 0.261 & {\cellcolor{yellow}} 0.824 & {\cellcolor{salmon}} 0.965 & {\cellcolor{salmon}} 0.0983 & {\cellcolor{yellow}} 0.797 & {\cellcolor{yellow}} 0.846 & {\cellcolor{lightgreen}} 0.671 & {\cellcolor{salmon}} 0.443 & {\cellcolor{salmon}} 0.245 & {\cellcolor{lightgreen}} 0.725 & {\cellcolor{lightgreen}} 0.631 & {\cellcolor{salmon}} 0.371 & {\cellcolor{lightgreen}} 0.703 & {\cellcolor{lightgreen}} 0.671 & {\cellcolor{lightgreen}} 0.649 \\
$f_{\text{ice}}$ & {\cellcolor{yellow}} 0.807 & {\cellcolor{lightgreen}} 0.695 & {\cellcolor{salmon}} 0.342 & {\cellcolor{salmon}} 0.885 & {\cellcolor{lightgreen}} 0.664 & {\cellcolor{lightgreen}} 0.682 & {\cellcolor{salmon}} 0.936 & {\cellcolor{lightgreen}} 0.628 & {\cellcolor{lightgreen}} 0.754 & {\cellcolor{yellow}} 0.861 & {\cellcolor{lightgreen}} 0.646 & {\cellcolor{salmon}} 0.477 & {\cellcolor{salmon}} 0.890 & {\cellcolor{lightgreen}} 0.664 & {\cellcolor{lightgreen}} 0.758 & {\cellcolor{yellow}} 0.876 & {\cellcolor{lightgreen}} 0.596 & {\cellcolor{lightgreen}} 0.782 \\
$r_{\text{effw}}$ & {\cellcolor{lightgreen}} 0.689 & {\cellcolor{lightgreen}} 0.685 & - & {\cellcolor{lightgreen}} 0.719 & {\cellcolor{lightgreen}} 0.736 & - & {\cellcolor{lightgreen}} 0.601 & {\cellcolor{yellow}} 0.785 & - & {\cellcolor{lightgreen}} 0.653 & {\cellcolor{lightgreen}} 0.649 & - & {\cellcolor{lightgreen}} 0.733 & {\cellcolor{lightgreen}} 0.683 & - & {\cellcolor{yellow}} 0.787 & {\cellcolor{lightgreen}} 0.767 & - \\
$r_{\text{effi}}$ & - & {\cellcolor{lightgreen}} 0.735 & {\cellcolor{yellow}} 0.524 & - & {\cellcolor{lightgreen}} 0.750 & {\cellcolor{lightgreen}} 0.693 & - & {\cellcolor{lightgreen}} 0.770 & {\cellcolor{lightgreen}} 0.766 & - & {\cellcolor{lightgreen}} 0.616 & {\cellcolor{yellow}} 0.490 & - & {\cellcolor{lightgreen}} 0.697 & {\cellcolor{lightgreen}} 0.671 & - & {\cellcolor{lightgreen}} 0.693 & {\cellcolor{lightgreen}} 0.739 \\
\bottomrule
\end{tabular}

\end{adjustbox}
\caption{1--$\sigma$ coverage for the set of defined situations (see Section~\ref{sec:situation_analysis} for the definitions). Green shading denote absolute departures from the nominal value 0.68 smaller than 0.1, yellow indicates absolute departures between 0.1 and 0.2, whereas red above 0.2. The fraction of points [\%] in the test dataset in each situation is reported in the first row (here, the cell shading is row-wise normalized and proportional to the cell value). For ice and water effective radii, "water" and "ice" situations are not shown, respectively, as the associated cloud water phase is absent.}
\label{tab:situations_coverage}
\end{table}

\subsection{Sensitivity to the Number and Type of Input Channels}\label{sec:sensitivity}

As shown in Fig.~\ref{fig:fg_correlation}, the MTG solar channels considered in this study are highly correlated. For this reason, it is important to test the contribution to the overall NN performance coming from each input channel or from a combination of them. In this section, we present a sensitivity study based on a series of models trained with different subsets of input solar channels and their corresponding albedos. Using multiple channels can help filter input noise and improve predictions. Here, however, we aim to assess each channel's intrinsic contribution to the sharpness of the retrieval. For this reason, we switch off input noise injection. In the following, model versions are named according to the set of input channels. We consider in total 9 model versions (namely, Only NIRs, NIR16-VIS6, NIRs-VIS4, NIRs-VIS5, NIRs-VIS6, NIRs-VIS8, NIRs-VIS4-VIS5, NIRs-VIS4-VIS5-VIS6, All), based on the multivariate Gaussian predictive distribution and trained with the energy score. All models except NIR16-VIS6 use both NIR channels, as their contribution to model performance was found to be substantial. We include NIR16-VIS6 as a reference, since 0.6$\mu m$ and 1.6$\mu m$ channels is one of the least correlated pairs, and, hence, the minimal and most informative set of channels in a DA perspective. The set NIRs-VISx including NIRs-VIS4, NIRs-VIS5, NIRs-VIS6, NIRs-VIS8 is meant to verify if any of the visible channels is more beneficial. In Table~\ref{tab:channel_comparison_RMSE_ES} the RMSE and ES are reported for all the model versions. Considering the set NIRs-VISx, no visible channel appears to have a higher impact on the metrics considered. This finding somewhat justifies the minimal sequence of models adopted to test the impact of including increasingly more channels. Considering such sequence (Only NIRs, NIRs-VIS4, NIRs-VIS4-VIS5, NIRs-VIS4-VIS5-VIS6, All), both metrics show that performance improves as more channels are included, converging to a $\approx20\%$ RMSE and ES reduction. As shown in Fig.~\ref{fig:fg_correlation}, visible channels show very high inter-channel correlations, with values approximately equal to 1, which limit the capabilities of a DA filter like LETKF to extract information from them. In contrast, the BO approach is not constrained by the assumptions associated to a typical DA filter (e.g., Gaussianity of the errors, linearity or quasi-linearity of the forward operator etc.), and in principle it can exploit the full non linear relationship between the model state $x$ (in our case simplified to BO's output space) and reflectances. For example, for two perfectly-correlated channels $R^{\lambda_1}$ and $R^{\lambda_2}$ for which Corr$[R^{\lambda_1}, R^{\lambda_2}]=1$, $R^{\lambda_1}$ can be expressed as $R^{\lambda_1}=\alpha R^{\lambda_2}+\beta$ for some coefficients $\alpha$ and $\beta$ that will in general depend on the regime, for example via the ensemble mean. The network may then still perform better by including both channels instead of only one, as it could in principle extract more information on the regime, which may come from $\alpha$ and $\beta$, or from the unconditional $(R^{\lambda_1}, R^{\lambda_2})$ distribution (marginalized with respect to $x$) that is available in the training data. In a more realistic scenario where the correlation is not perfect yet very high, the BO approach could leverage more also on the small departures from linearity compared to the assimilation with LETKF, since the BO does not need to include the representativeness component in the input errors.

\begin{table}
\centering
\begin{adjustbox}{max width=\textwidth}
    \begin{tabular}{l|ccccccccc}
\toprule
 & OnlyNIRs & NIR16-VIS6 & NIRs-VIS4 & NIRs-VIS5 & NIRs-VIS6 & NIRs-VIS8 & NIRs-VIS4-VIS5 & NIRs-VIS4-VIS5-VIS6 & All \\
\midrule
$\log(\tau)$  & {\cellcolor{lightblue!100}} 0.5435 & {\cellcolor{lightblue!56}} 0.4674 & {\cellcolor{lightblue!39}} 0.4392 & {\cellcolor{lightblue!39}} 0.4390 & {\cellcolor{lightblue!41}} 0.4415 & {\cellcolor{lightblue!39}} 0.4396 & {\cellcolor{lightblue!26}} 0.4159 & {\cellcolor{lightblue!20}} 0.4051 & {\cellcolor{lightblue!15}} 0.3965 \\
$\gamma$  & {\cellcolor{lightblue!100}} 0.1259 & {\cellcolor{lightblue!89}} 0.1226 & {\cellcolor{lightblue!80}} 0.1198 & {\cellcolor{lightblue!72}} 0.1174 & {\cellcolor{lightblue!74}} 0.1181 & {\cellcolor{lightblue!72}} 0.1173 & {\cellcolor{lightblue!33}} 0.1054 & {\cellcolor{lightblue!23}} 0.1022 & {\cellcolor{lightblue!15}} 0.09968 \\
$f_{\text{ice}}$  & {\cellcolor{lightblue!100}} 0.1200 & {\cellcolor{lightblue!82}} 0.1147 & {\cellcolor{lightblue!73}} 0.1119 & {\cellcolor{lightblue!69}} 0.1106 & {\cellcolor{lightblue!70}} 0.1109 & {\cellcolor{lightblue!50}} 0.1046 & {\cellcolor{lightblue!33}} 0.09925 & {\cellcolor{lightblue!27}} 0.09740 & {\cellcolor{lightblue!15}} 0.09347 \\
$r_{\text{effw}}$ [$\mu m$] & {\cellcolor{lightblue!100}} 1.372 & {\cellcolor{lightblue!72}} 1.249 & {\cellcolor{lightblue!41}} 1.109 & {\cellcolor{lightblue!40}} 1.105 & {\cellcolor{lightblue!42}} 1.116 & {\cellcolor{lightblue!46}} 1.135 & {\cellcolor{lightblue!31}} 1.068 & {\cellcolor{lightblue!21}} 1.021 & {\cellcolor{lightblue!15}} 0.9929 \\
$r_{\text{effi}}$ [$\mu m$] & {\cellcolor{lightblue!100}} 7.510 & {\cellcolor{lightblue!71}} 7.169 & {\cellcolor{lightblue!64}} 7.088 & {\cellcolor{lightblue!58}} 7.012 & {\cellcolor{lightblue!62}} 7.056 & {\cellcolor{lightblue!45}} 6.856 & {\cellcolor{lightblue!36}} 6.754 & {\cellcolor{lightblue!27}} 6.640 & {\cellcolor{lightblue!15}} 6.490 \\
\midrule
ES & {\cellcolor{lightblue!100}} 3.230 & {\cellcolor{lightblue!73}} 3.012 & {\cellcolor{lightblue!66}} 2.947 & {\cellcolor{lightblue!60}} 2.900 & {\cellcolor{lightblue!63}} 2.926 & {\cellcolor{lightblue!50}} 2.818 & {\cellcolor{lightblue!37}} 2.708 & {\cellcolor{lightblue!26}} 2.616 & {\cellcolor{lightblue!15}} 2.521 \\
\bottomrule
\end{tabular}

\end{adjustbox}
\caption{Root Mean Squared Error (RMSE) of the output variables and Energy Score (ES) for the set of models relevant to assess the sensitivity to input channels (see Section~\ref{sec:sensitivity}). The cell shading is row-wise normalized, and proportional to the cell value.}
\label{tab:channel_comparison_RMSE_ES}
\end{table}

\section{Conclusions and Outlook}\label{section:conclusion_outlook}

In this work, the joint information content of observations in six solar channels of the FCI instrument onboard MTG was assessed. For this purpose, we introduced a "Backward Operator" (BO) based on a Distributional Regression Network (DRN). The proposed method provides probabilistic retrievals of cloud variables whose conditional distribution is assumed to follow a multivariate Gaussian distribution. The cloud variables retrieved by the BO are the total optical thickness, the column cloud fraction, the ice fraction, and the effective radii for water and ice cloud particles. Uncertainties in the reflectances and albedos of the six channels, which are inputs of the BO, were considered during the training phase by injecting noise according to a prescribed heteroscedastic error model. The BO avoids typical drawbacks of common retrieval methods in the context of DA. Specifically, it does not require prior information, it guarantees the consistency between the retrieval method and the DA system, as they are based on the same forward operator, and it provides a situation-dependent uncertainty covariance matrix for the retrieved quantities. Globally, the BO showed negligible bias and good calibration. For the predicted uncertainties, a good coverage was found at the 1--$\sigma$ level which indicates that the estimates are reliable. Regarding the predicted covariances, a non-trivial covariance structure was found, in accordance with the ground truth provided by a sampling method. The analysis of information content showed that optical thickness and column cloud fraction are best constrained in optically thick and fully cloudy conditions, while ice fraction uncertainty mainly depends on the ice fraction itself, with smallest uncertainties in only-water situations. The uncertainty of the effective radii depends in a more complex way on the other variables, with relatively low values found for thin clouds and very thick clouds and increased uncertainty for partially cloudy cases at intermediate optical depths. Finally, the sensitivity of the retrievals performance to the number and type of input channels was assessed by training many BO versions using different subsets of the solar channels considered, assuming perfectly-observed inputs. The combination of the 1.6$\mu m$ near-infrared channel and the 0.6$\mu m$ visible channel outperformed the combination of two near-infrared channels. All combinations of the two near-infrared channels and any of the visible channels showed similar performance. However, including more visible channels still resulted in significantly improved performance (around 20\% reduction in RMSE and ES), although these channels are highly correlated. A similar approach to the one presented here could also prove useful in designing new satellite instruments.

The current implementation of the BO has some limitations. For example, the representation of subgrid cloud overlap in the forward operator is not yet fully consistent with the NWP model, and three-dimensional radiative transfer effects are not taken into account. Moreover, the representation of clouds in the NWP model cannot be expected to be perfect, and therefore unrealistic model states may be contained in the training data set, and realistic ones may be missing. These limitations affect the fidelity of the retrievals, though it is worth noting that they would equally affect direct radiance assimilation approaches that rely on the same forward operator. In addition, the representation of climatology is limited, as it includes only summer-like scenes over the ICON-D2 limited area domain over Germany. Future work should aim to address these shortcomings, for example, by including an improved forward operator, by improving the representation of clouds in the NWP model, or simply by extending the training data set to include a broader range of cloud regimes and geographic regions. The long-term goal, which is beyond the scope of this proof-of-concept study, would be to apply our method to real observations and to compare the retrievals to products obtained with established retrieval algorithms.

Another important step will be to assimilate the BO retrievals and to assess their impact relative to direct radiance assimilation. For the assimilation of retrievals, the situation-dependent covariance matrix generated by the BO could be used as the error covariance matrix, together with a component related to the representation error. This assimilation strategy is promising for several reasons: retrieved quantities typically exhibit lower inter-channel correlations than reflectances, are more directly related to model state variables, and thus have the potential to mitigate both ambiguity and non-linearity errors. Moreover, assimilating BO retrievals may allow us to fully exploit the information contained in visible channels despite their very high correlation, which could be challenging for direct assimilation. Retrieval assimilation would also offer practical efficiency advantages, as the BO needs to be called only once per analysis time, condensing the information from many channels into a reduced set of quantities. In contrast, direct assimilation requires the forward operator to be run for every channel and every ensemble member.

\section*{Acknowledgments}

This study was funded by the Hans Ertel Centre for Weather Research. This German research network of universities and research institutes as well as DWD is funded by the Federal Ministry for Digital and Transport (BMDV, grant no. DWD2014P8).

\clearpage
\appendix
\renewcommand{\appendixname}{Appendix}
\section{Hyperparameter Tuning}

In this section, we describe the final model, as well as the corresponding hyperparameter and architecture tuning framework.
As mentioned earlier, all experiments were performed using MLPs with several hidden layers, as those are most suitable for this prediction task. However, model performance strongly depends on both the network architecture and the chosen training regime. While exploring the whole space of possible hyperparameters is computationally infeasible, approximation algorithms can be used to optimize the hyperparameters and obtain an approximation of the ``best'' model. For that purpose, we run extensive hyper-parameter tuning using the Ray Tune library \citep{liaw2018tuneresearchplatformdistributed} and the asynchronous successive halving algorithm \citep{li2020massivelyparallelhyperparametertuning}. In particular, we sample 250 randomly sampled parameter configurations from the corresponding search spaces in Table~\ref{tab:hparam_configuration}, train the model based on energy score for 250 epochs, and evaluate the corresponding validation loss. The resulting five best hyper-parameter configurations are shown in Table~\ref{tab:hparam_results}, with the best and chosen model highlighted. The results show that the number of neurons seems to be consistent across the best-performing runs, i.e., $n_\mathrm{neurons} = 256$. Similarly, the learning rate and dropout are all within the same order of magnitude, showing that the results are somewhat robust with regard to the best model. However, for the batch size and especially the activation function, there seems to be no direct correspondence to the model performance. While the relationship between the different hyperparameters and the model performance could be further studied, this is beyond the scope of this article.

\begin{table}
\centering
\begin{adjustbox}{max width=\textwidth}
    \begin{tabular}{l|rr}
\toprule
Parameter & Sampling method & Sampling domain \\
\midrule
Learning rate &Log-uniform & $[5 \times 10^{-5}, 1\times 10^{-1}]$ \\
Batch size & Uniform& $ 2^s, s\in\{10,11,12,13,14 \}$\\
Number of hidden layers & Uniform &$\{2,4,6,8\}$ \\
Neurons per hidden layer & Uniform &$2^s, s \in\{5,6,7,8\}$ \\
Dropout &Uniform & $(0,0.1]$\\
Activation function & Uniform&$\{\mathrm{ReLU}, \mathrm{ELU}, \mathrm{GELU}, \mathrm{SiLU}, \mathrm{Softplus}, \mathrm{Sigmoid}\}$ \\

\bottomrule
\end{tabular}

\end{adjustbox}
\caption{Search space for the hyperparameters of the neural network and training regime.}
\label{tab:hparam_configuration}
\end{table}

\begin{table}
\centering
\begin{adjustbox}{max width=\textwidth}
    \begin{tabular}{l|rrrrrrr}
\toprule
Rank &Loss & Learning rate & Batch size & $n_\mathrm{hidden}$ & $n_\mathrm{neurons}$& Dropout & Activation \\
\midrule
\textbf{1} & \textbf{0.1415} & $\mathbf{3.125 \times 10 ^{-3}}$ & \textbf{8192} & \textbf{8} & \textbf{256} &$\mathbf{1.133\times 10 ^{-3}}$ & \textbf{GELU}\\
2 & 0.1536 & $1.642\times 10^{-3}$& 4096 &6 &256 &${4.449\times 10 ^{-3}}$& ELU\\
3& 0.1614 & $4.155\times 10^{-3}$ & 1024 &6&256 &${5.463\times 10 ^{-2}}$ & Sigmoid\\
4 & 0.1621 &$0.500\times 10^{-3}$ & 2048 &8&256 &${3.958\times 10 ^{-2}}$ &Softplus\\
5 & 0.1624 &$0.745\times 10^{-3}$& 1024 &4 &256 &${8.347\times 10 ^{-2}}$& GELU\\
\bottomrule
\end{tabular}

\end{adjustbox}
\caption{The five hyper-parameter configurations with the lowest validation loss (energy score for the normalised output variables), with the best model highlighted in bold.}
\label{tab:hparam_results}
\end{table}

\bibliographystyle{abbrvnat}
\bibliography{bibliography}

@article {rasp2018,
      author = "Stephan Rasp and Sebastian Lerch",
      title = "Neural Networks for Postprocessing Ensemble Weather Forecasts",
      journal = "Monthly Weather Review",
      year = "2018",
      publisher = "American Meteorological Society",
      address = "Boston MA, USA",
      volume = "146",
      number = "11",
      doi = "10.1175/MWR-D-18-0187.1",
      pages=      "3885 - 3900",
      url = "https://journals.ametsoc.org/view/journals/mwre/146/11/mwr-d-18-0187.1.xml"
}

@article{gneiting2007scoringrules,
author = {Gneiting, Tilmann and Raftery, Adrian},
year = {2007},
month = {03},
pages = {359-378},
title = {Strictly Proper Scoring Rules, Prediction, and Estimation},
volume = {102},
journal = {Journal of the American Statistical Association},
doi = {10.1198/016214506000001437}
}

@article {chapman2022,
      author = "William E. Chapman and Luca Delle Monache and Stefano Alessandrini and Aneesh C. Subramanian and F. Martin Ralph and Shang-Ping Xie and Sebastian Lerch and Negin Hayatbini",
      title = "Probabilistic Predictions from Deterministic Atmospheric River Forecasts with Deep Learning",
      journal = "Monthly Weather Review",
      year = "2022",
      publisher = "American Meteorological Society",
      address = "Boston MA, USA",
      volume = "150",
      number = "1",
      doi = "10.1175/MWR-D-21-0106.1",
      pages=      "215 - 234",
      url = "https://journals.ametsoc.org/view/journals/mwre/150/1/MWR-D-21-0106.1.xml"
}

@article {schulz2022,
      author = "Benedikt Schulz and Sebastian Lerch",
      title = "Machine Learning Methods for Postprocessing Ensemble Forecasts of Wind Gusts: A Systematic Comparison",
      journal = "Monthly Weather Review",
      year = "2022",
      publisher = "American Meteorological Society",
      address = "Boston MA, USA",
      volume = "150",
      number = "1",
      doi = "10.1175/MWR-D-21-0150.1",
      pages=      "235 - 257",
      url = "https://journals.ametsoc.org/view/journals/mwre/150/1/MWR-D-21-0150.1.xml"
}

@article {gneiting2005,
      author = "Tilmann Gneiting and Adrian E. Raftery and Anton H. Westveld and Tom Goldman",
      title = "Calibrated Probabilistic Forecasting Using Ensemble Model Output Statistics and Minimum CRPS Estimation",
      journal = "Monthly Weather Review",
      year = "2005",
      publisher = "American Meteorological Society",
      address = "Boston MA, USA",
      volume = "133",
      number = "5",
      doi = "10.1175/MWR2904.1",
      pages=      "1098 - 1118",
      url = "https://journals.ametsoc.org/view/journals/mwre/133/5/mwr2904.1.xml"
}

@article{gneiting2007calibrationsharpness,
author = {Gneiting, Tilmann and Balabdaoui, Fadoua and Raftery, Adrian E.},
title = {Probabilistic forecasts, calibration and sharpness},
journal = {Journal of the Royal Statistical Society: Series B (Statistical Methodology)},
volume = {69},
year = {2007},
number = {2},
pages = {243-268},
doi = {https://doi.org/10.1111/j.1467-9868.2007.00587.x},
url = {https://rss.onlinelibrary.wiley.com/doi/abs/10.1111/j.1467-9868.2007.00587.x},
eprint = {https://rss.onlinelibrary.wiley.com/doi/pdf/10.1111/j.1467-9868.2007.00587.x},
year = {2007}
}

@article{neyman1937,
    author = {Neyman, J.},
    title = {Outline of a Theory of Statistical Estimation Based on the Classical Theory of Probability},
    journal = {Philosophical Transactions of the Royal Society of London, Series A: Mathematical and Physical Sciences},
    volume = {236},
    number = {767},
    pages = {333-380},
    year = {1937},
    month = {08},
    issn = {0080-4614},
    doi = {10.1098/rsta.1937.0005},
    url = {https://doi.org/10.1098/rsta.1937.0005},
    eprint = {https://royalsocietypublishing.org/rsta/article-pdf/236/767/333/47684/rsta.1937.0005.pdf},
}

@article{hyndman1996,
 ISSN = {10618600},
 URL = {http://www.jstor.org/stable/1390887},
 author = {Rob J. Hyndman and David M. Bashtannyk and Gary K. Grunwald},
 journal = {Journal of Computational and Graphical Statistics},
 number = {4},
 pages = {315--336},
 publisher = {[American Statistical Association, Taylor & Francis, Ltd., Institute of Mathematical Statistics, Interface Foundation of America]},
 title = {Estimating and Visualizing Conditional Densities},
 urldate = {2026-01-25},
 volume = {5},
 year = {1996}
}

@article{kugler2025,
author = {Kugler, Lukas and Weissmann, Martin},
title = {Effects of Observation-Operator Nonlinearity on the Assimilation of Visible and Infrared Radiances in Ensemble Data Assimilation},
journal = {Quarterly Journal of the Royal Meteorological Society},
volume = {151},
number = {770},
pages = {e4970},
doi = {https://doi.org/10.1002/qj.4970},
url = {https://rmets.onlinelibrary.wiley.com/doi/abs/10.1002/qj.4970},
eprint = {https://rmets.onlinelibrary.wiley.com/doi/pdf/10.1002/qj.4970},
year = {2025}
}

@article{scheck2016,
title = {A fast radiative transfer method for the simulation of visible satellite imagery},
journal = {Journal of Quantitative Spectroscopy and Radiative Transfer},
volume = {175},
pages = {54-67},
year = {2016},
issn = {0022-4073},
doi = {https://doi.org/10.1016/j.jqsrt.2016.02.008},
url = {https://www.sciencedirect.com/science/article/pii/S0022407316000285},
author = {Leonhard Scheck and Pascal Frèrebeau and Robert Buras-Schnell and Bernhard Mayer}
}

@article{scheck2020,
author = {Scheck, Leonhard and Weissmann, Martin and Bach, Liselotte},
title = {Assimilating visible satellite images for convective-scale numerical weather prediction: A case-study},
journal = {Quarterly Journal of the Royal Meteorological Society},
volume = {146},
number = {732},
pages = {3165-3186},
doi = {https://doi.org/10.1002/qj.3840},
url = {https://rmets.onlinelibrary.wiley.com/doi/abs/10.1002/qj.3840},
eprint = {https://rmets.onlinelibrary.wiley.com/doi/pdf/10.1002/qj.3840},
year = {2020}
}

@Article{baur2023,
AUTHOR = {Baur, F. and Scheck, L. and Stumpf, C. and K\"opken-Watts, C. and Potthast, R.},
TITLE = {A neural-network-based method for generating synthetic 1.6\,$\mu m$ near-infrared satellite images},
JOURNAL = {Atmospheric Measurement Techniques},
VOLUME = {16},
YEAR = {2023},
NUMBER = {21},
PAGES = {5305--5326},
URL = {https://amt.copernicus.org/articles/16/5305/2023/},
DOI = {10.5194/amt-16-5305-2023}
}

@Article{geer2019,
AUTHOR = {Geer, A. J. and Migliorini, S. and Matricardi, M.},
TITLE = {All-sky assimilation of infrared radiances sensitive to mid- and upper-tropospheric moisture and cloud},
JOURNAL = {Atmospheric Measurement Techniques},
VOLUME = {12},
YEAR = {2019},
NUMBER = {9},
PAGES = {4903--4929},
URL = {https://amt.copernicus.org/articles/12/4903/2019/},
DOI = {10.5194/amt-12-4903-2019}
}

@article{geer2018,
author = {Geer, Alan J. and Lonitz, Katrin and Weston, Peter and Kazumori, Masahiro and Okamoto, Kozo and Zhu, Yanqiu and Liu, Emily Huichun and Collard, Andrew and Bell, William and Migliorini, Stefano and Chambon, Philippe and Fourrié, Nadia and Kim, Min-Jeong and Köpken-Watts, Christina and Schraff, Christoph},
title = {All-sky satellite data assimilation at operational weather forecasting centres},
journal = {Quarterly Journal of the Royal Meteorological Society},
volume = {144},
number = {713},
pages = {1191-1217},
doi = {https://doi.org/10.1002/qj.3202},
url = {https://rmets.onlinelibrary.wiley.com/doi/abs/10.1002/qj.3202},
eprint = {https://rmets.onlinelibrary.wiley.com/doi/pdf/10.1002/qj.3202},
year = {2018}
}

@article { nakajima1990,
      author = "Teruyuki  Nakajima and Michael D.  King",
      title = "Determination of the Optical Thickness and Effective Particle Radius of Clouds from Reflected Solar Radiation Measurements. Part I: Theory",
      journal = "Journal of Atmospheric Sciences",
      year = "1990",
      publisher = "American Meteorological Society",
      address = "Boston MA, USA",
      volume = "47",
      number = "15",
      doi = "10.1175/1520-0469(1990)047<1878:DOTOTA>2.0.CO;2",
      pages=      "1878 - 1893",
      url = "https://journals.ametsoc.org/view/journals/atsc/47/15/1520-0469_1990_047_1878_dotota_2_0_co_2.xml"
}

@article{wattsOCA2011,
author = {Watts, P. D. and Bennartz, R. and Fell, F.},
title = {Retrieval of two-layer cloud properties from multispectral observations using optimal estimation},
journal = {Journal of Geophysical Research: Atmospheres},
volume = {116},
number = {D16},
pages = {},
doi = {https://doi.org/10.1029/2011JD015883},
url = {https://agupubs.onlinelibrary.wiley.com/doi/abs/10.1029/2011JD015883},
eprint = {https://agupubs.onlinelibrary.wiley.com/doi/pdf/10.1029/2011JD015883},
year = {2011}
}

@ARTICLE{gonzalez2025,
  author={Gonzalez, Jessenia and Dipu, Sudhakar and Jimenez, Gabriel and Camps-Valls, Gustau and Quaas, Johannes},
  journal={IEEE Journal of Selected Topics in Applied Earth Observations and Remote Sensing}, 
  title={Machine Learning-Based Retrieval of Cloud Droplet Number Concentration and Liquid Water Path From Satellite Spectral Data}, 
  year={2025},
  volume={18},
  number={},
  pages={21910-21922},
  doi={10.1109/JSTARS.2025.3601981}}

@Article{chen2022,
AUTHOR = {Chen, Xingfeng and Zhao, Limin and Zheng, Fengjie and Li, Jiaguo and Li, Lei and Ding, Haonan and Zhang, Kainan and Liu, Shumin and Li, Donghui and de Leeuw, Gerrit},
TITLE = {Neural Network AEROsol Retrieval for Geostationary Satellite (NNAeroG) Based on Temporal, Spatial and Spectral Measurements},
JOURNAL = {Remote Sensing},
VOLUME = {14},
YEAR = {2022},
NUMBER = {4},
ARTICLE-NUMBER = {980},
URL = {https://www.mdpi.com/2072-4292/14/4/980},
ISSN = {2072-4292},
DOI = {10.3390/rs14040980}
}

@article{zangl2015,
author = {Zängl, Günther and Reinert, Daniel and Rípodas, Pilar and Baldauf, Michael},
title = {The ICON (ICOsahedral Non-hydrostatic) modelling framework of DWD and MPI-M: Description of the non-hydrostatic dynamical core},
journal = {Quarterly Journal of the Royal Meteorological Society},
volume = {141},
number = {687},
pages = {563-579},
doi = {https://doi.org/10.1002/qj.2378},
url = {https://rmets.onlinelibrary.wiley.com/doi/abs/10.1002/qj.2378},
eprint = {https://rmets.onlinelibrary.wiley.com/doi/pdf/10.1002/qj.2378},
year = {2015}
}

@Article{saunders2018rttovv13,
AUTHOR = {Saunders, R. and Hocking, J. and Turner, E. and Rayer, P. and Rundle, D. and Brunel, P. and Vidot, J. and Roquet, P. and Matricardi, M. and Geer, A. and Bormann, N. and Lupu, C.},
TITLE = {An update on the RTTOV fast radiative transfer model (currently at version~12)},
JOURNAL = {Geoscientific Model Development},
VOLUME = {11},
YEAR = {2018},
NUMBER = {7},
PAGES = {2717--2737},
URL = {https://gmd.copernicus.org/articles/11/2717/2018/},
DOI = {10.5194/gmd-11-2717-2018}
}

@Article{kurzrock2018,
author = "Kurzrock, Frederik and Cros, Sylvain and Ming, Fabrice Chane and Otkin, Jason A. and Hutt, Axel and Linguet, Laurent and Lajoie, Gilles and Potthast, Roland",
journal = "Meteorologische Zeitschrift",
month = 11,
year = 2018,
title = "A Review of the Use of Geostationary Satellite Observations in Regional-Scale Models for Short-term Cloud Forecasting",
number = "4",
volume = "27",
pages = {277-298},
url = "http://dx.doi.org/10.1127/metz/2018/0904",
doi = "10.1127/metz/2018/0904",
publisher = "Schweizerbart Science Publishers",
address = "Stuttgart, Germany"
}

@misc{bormann2019,
  author = {Niels Bormann and Heather Lawrence and Jacqueline Farnan},
  title = {Global observing system experiments in the ECMWF assimilation system},
  year = {2019},
  journal = {ECMWF Technical Memoranda},
  number = {839},
  publisher = {ECMWF},
  url = {https://www.ecmwf.int/node/18859},
  doi = {10.21957/sr184iyz},
  language = {eng},
}

@misc{FCI_val_report,
  title     = {FCI L1 Operational Product Report},
  author    = {{EUMETSAT}},
  year      = {2024},
  month     = {December},
  day       = {10},
  institution = {EUMETSAT},
  note      = {Retrieved from \url{https://user.eumetsat.int/s3/eup-strapi-media/FCI\_L1\_Operational\_Product\_Validation\_Report\_eb9e6db47c.pdf}}
}

@Article{BRDFatlas,
AUTHOR = {Vidot, Jérôme and Brunel, Pascal and Dumont, Marie and Carmagnola, Carlo and Hocking, James},
TITLE = {The VIS/NIR Land and Snow BRDF Atlas for RTTOV: Comparison between MODIS MCD43C1 C5 and C6},
JOURNAL = {Remote Sensing},
VOLUME = {10},
YEAR = {2018},
NUMBER = {1},
ARTICLE-NUMBER = {21},
URL = {https://www.mdpi.com/2072-4292/10/1/21},
ISSN = {2072-4292},
DOI = {10.3390/rs10010021}
}

@article{eyre2022,
author = {Eyre, J. R. and Bell, W. and Cotton, J. and English, S. J. and Forsythe, M. and Healy, S. B. and Pavelin, E. G.},
title = {Assimilation of satellite data in numerical weather prediction. Part II: Recent years},
journal = {Quarterly Journal of the Royal Meteorological Society},
volume = {148},
number = {743},
pages = {521-556},
doi = {https://doi.org/10.1002/qj.4228},
url = {https://rmets.onlinelibrary.wiley.com/doi/abs/10.1002/qj.4228},
eprint = {https://rmets.onlinelibrary.wiley.com/doi/pdf/10.1002/qj.4228},
year = {2022}
}

@misc{liaw2018tuneresearchplatformdistributed,
      title={Tune: A Research Platform for Distributed Model Selection and Training}, 
      author={Richard Liaw and Eric Liang and Robert Nishihara and Philipp Moritz and Joseph E. Gonzalez and Ion Stoica},
      year={2018},
      eprint={1807.05118},
      archivePrefix={arXiv},
      primaryClass={cs.LG},
      url={https://arxiv.org/abs/1807.05118}, 
}

@article{bultepno,
  title = {Probabilistic neural operators for functional uncertainty quantification},
  author = {B{\"u}lte, Christopher and Scholl, Philipp and Kutyniok, Gitta},
  journal = {Transactions on Machine Learning Research},
  issn = {2835-8856},
  year = {2025},
  url = {https://openreview.net/forum?id=gangoPXSRw},
  note = {},
}

@misc{li2020massivelyparallelhyperparametertuning,
      title={A System for Massively Parallel Hyperparameter Tuning}, 
      author={Liam Li and Kevin Jamieson and Afshin Rostamizadeh and Ekaterina Gonina and Moritz Hardt and Benjamin Recht and Ameet Talwalkar},
      year={2020},
      eprint={1810.05934},
      archivePrefix={arXiv},
      primaryClass={cs.LG},
      url={https://arxiv.org/abs/1810.05934}, 
}

@article{muschinski,
  title = {Cholesky-Based Multivariate {{Gaussian}} Regression},
  author = {Muschinski, Thomas and Mayr, Georg J. and Simon, Thorsten and Umlauf, Nikolaus and Zeileis, Achim},
  year = 2024,
  month = jan,
  journal = {Econometrics and Statistics},
  volume = {29},
  pages = {261--281},
  issn = {2452-3062},
  doi = {10.1016/j.ecosta.2022.03.001},
  urldate = {2025-06-04}
}

@misc{buchweitz2025asymmetricpenaltiesunderlieproper,
      title={Asymmetric Penalties Underlie Proper Loss Functions in Probabilistic Forecasting}, 
      author={Erez Buchweitz and João Vitor Romano and Ryan J. Tibshirani},
      year={2025},
      eprint={2505.00937},
      archivePrefix={arXiv},
      primaryClass={math.ST},
      url={https://arxiv.org/abs/2505.00937}, 
}

@inproceedings{NEURIPS2019_07211688,
 author = {Skafte, Nicki and J\o rgensen, Martin and Hauberg, S\o ren},
 booktitle = {Advances in Neural Information Processing Systems},
 editor = {H. Wallach and H. Larochelle and A. Beygelzimer and F. d\textquotesingle Alch\'{e}-Buc and E. Fox and R. Garnett},
 pages = {},
 publisher = {Curran Associates, Inc.},
 title = {Reliable training and estimation of variance networks},
 url = {https://proceedings.neurips.cc/paper_files/paper/2019/file/07211688a0869d995947a8fb11b215d6-Paper.pdf},
 volume = {32},
 year = {2019}
}

@inproceedings{NEURIPS2023_a901d554,
 author = {Immer, Alexander and Palumbo, Emanuele and Marx, Alexander and Vogt, Julia},
 booktitle = {Advances in Neural Information Processing Systems},
 editor = {A. Oh and T. Naumann and A. Globerson and K. Saenko and M. Hardt and S. Levine},
 pages = {53996--54019},
 publisher = {Curran Associates, Inc.},
 title = {Effective Bayesian Heteroscedastic Regression with Deep Neural Networks},
 url = {https://proceedings.neurips.cc/paper_files/paper/2023/file/a901d5540789a086ee0881a82211b63d-Paper-Conference.pdf},
 volume = {36},
 year = {2023}
}

@article{pacchiardi,
  title = {Probabilistic Forecasting with Generative Networks via Scoring Rule Minimization},
  author = {Pacchiardi, Lorenzo and Adewoyin, Rilwan and Dueben, Peter and Dutta, Ritabrata},
  year = 2021
}

@article{engression,
  title = {Engression: Extrapolation through the Lens of Distributional Regression},
  shorttitle = {Engression},
  author = {Shen, Xinwei and Meinshausen, Nicolai},
  year = 2025,
  month = jul,
  journal = {Journal of the Royal Statistical Society Series B: Statistical Methodology},
  volume = {87},
  number = {3},
  pages = {653--677},
  issn = {1369-7412},
  doi = {10.1093/jrsssb/qkae108},
  urldate = {2025-09-01}
}

@article{chen_generative,
  title = {Generative Machine Learning Methods for Multivariate Ensemble Postprocessing},
  author = {Chen, Jieyu and Janke, Tim and Steinke, Florian and Lerch, Sebastian},
  year = 2024,
  month = mar,
  journal = {The Annals of Applied Statistics},
  volume = {18},
  number = {1},
  pages = {159--183},
  publisher = {Institute of Mathematical Statistics},
  issn = {1932-6157, 1941-7330},
  doi = {10.1214/23-AOAS1784},
  urldate = {2026-02-20},
  langid = {english}
}

@article{MAYER2026114361,
title = {Post-processing of ensemble photovoltaic power forecasts with distributional and quantile regression methods},
journal = {Solar Energy},
volume = {307},
pages = {114361},
year = {2026},
issn = {0038-092X},
doi = {https://doi.org/10.1016/j.solener.2026.114361},
url = {https://www.sciencedirect.com/science/article/pii/S0038092X26000496},
author = {Martin János Mayer and Ágnes Baran and Sebastian Lerch and Nina Horat and Dazhi Yang and Sándor Baran}
}

@misc{chen2025probabilisticintradayelectricityprice,
      title={Probabilistic intraday electricity price forecasting using generative machine learning}, 
      author={Jieyu Chen and Sebastian Lerch and Melanie Schienle and Tomasz Serafin and Rafał Weron},
      year={2025},
      eprint={2506.00044},
      archivePrefix={arXiv},
      primaryClass={stat.AP},
      url={https://arxiv.org/abs/2506.00044}, 
}

@article{szekely_test,
  title = {A New Test for Multivariate Normality},
  author = {Sz{\'e}kely, G{\'a}bor J. and Rizzo, Maria L.},
  year = 2005,
  month = mar,
  journal = {Journal of Multivariate Analysis},
  volume = {93},
  number = {1},
  pages = {58--80},
  issn = {0047-259X},
  doi = {10.1016/j.jmva.2003.12.002},
  urldate = {2024-03-22}
}

@article{KNEIB202399,
title = {Rage Against the Mean – A Review of Distributional Regression Approaches},
journal = {Econometrics and Statistics},
volume = {26},
pages = {99-123},
year = {2023},
issn = {2452-3062},
doi = {https://doi.org/10.1016/j.ecosta.2021.07.006},
url = {https://www.sciencedirect.com/science/article/pii/S2452306221000824},
author = {Thomas Kneib and Alexander Silbersdorff and Benjamin Säfken}
}

@article { rossow1999,
      author = "William B. Rossow and Robert A. Schiffer",
      title = "Advances in Understanding Clouds from ISCCP",
      journal = "Bulletin of the American Meteorological Society",
      year = "1999",
      publisher = "American Meteorological Society",
      address = "Boston MA, USA",
      volume = "80",
      number = "11",
      doi = "10.1175/1520-0477(1999)080<2261:AIUCFI>2.0.CO;2",
      pages=      "2261 - 2288",
      url = "https://journals.ametsoc.org/view/journals/bams/80/11/1520-0477_1999_080_2261_aiucfi_2_0_co_2.xml"
}

@article{pincus2008,
author = {Pincus, Robert and Batstone, Crispian P. and Hofmann, Robert J. Patrick and Taylor, Karl E. and Glecker, Peter J.},
title = {Evaluating the present-day simulation of clouds, precipitation, and radiation in climate models},
journal = {Journal of Geophysical Research: Atmospheres},
volume = {113},
number = {D14},
pages = {},
doi = {https://doi.org/10.1029/2007JD009334},
url = {https://agupubs.onlinelibrary.wiley.com/doi/abs/10.1029/2007JD009334},
eprint = {https://agupubs.onlinelibrary.wiley.com/doi/pdf/10.1029/2007JD009334},
year = {2008}
}

@article { stephens2005,
      author = "Graeme L. Stephens",
      title = "Cloud Feedbacks in the Climate System: A Critical Review",
      journal = "Journal of Climate",
      year = "2005",
      publisher = "American Meteorological Society",
      address = "Boston MA, USA",
      volume = "18",
      number = "2",
      doi = "10.1175/JCLI-3243.1",
      pages=      "237 - 273",
      url = "https://journals.ametsoc.org/view/journals/clim/18/2/jcli-3243.1.xml"
}

@article{eyre1993assimilation,
  title={Assimilation of TOVS radiance information through one-dimensional variational analysis},
  author={Eyre, JR and Kelly, GA and McNally, AP and Andersson, E and Persson, A},
  journal={Quarterly Journal of the Royal Meteorological Society},
  volume={119},
  number={514},
  pages={1427--1463},
  year={1993},
  publisher={Wiley Online Library}
}

@article{bouttier2001,
author = {Bouttier, F. and Kelly, G.},
title = {Observing-system experiments in the ECMWF 4D-Var data assimilation system},
journal = {Quarterly Journal of the Royal Meteorological Society},
volume = {127},
number = {574},
pages = {1469-1488},
doi = {https://doi.org/10.1002/qj.49712757419},
url = {https://rmets.onlinelibrary.wiley.com/doi/abs/10.1002/qj.49712757419},
eprint = {https://rmets.onlinelibrary.wiley.com/doi/pdf/10.1002/qj.49712757419},
year = {2001}
}

@inproceedings{rodgers2000,
  title={Inverse Methods for Atmospheric Sounding: Theory and Practice},
  author={Clive D. Rodgers},
  year={2000},
  url={https://api.semanticscholar.org/CorpusID:60696486}
}

@article{hotta2021,
   title={Why does EnKF suffer from analysis overconfidence? An insight into exploiting the ever‐increasing volume of observations},
   volume={147},
   ISSN={1477-870X},
   url={http://dx.doi.org/10.1002/qj.3970},
   DOI={10.1002/qj.3970},
   number={735},
   journal={Quarterly Journal of the Royal Meteorological Society},
   publisher={Wiley},
   author={Hotta, Daisuke and Ota, Yoichiro},
   year={2021},
   month=jan, pages={1258–1277} }

@article{desroziers2005,
author = {Desroziers, G. and Berre, L. and Chapnik, B. and Poli, P.},
title = {Diagnosis of observation, background and analysis-error statistics in observation space},
journal = {Quarterly Journal of the Royal Meteorological Society},
volume = {131},
number = {613},
pages = {3385-3396},
doi = {https://doi.org/10.1256/qj.05.108},
url = {https://rmets.onlinelibrary.wiley.com/doi/abs/10.1256/qj.05.108},
eprint = {https://rmets.onlinelibrary.wiley.com/doi/pdf/10.1256/qj.05.108},
year = {2005}
}

@article { Holmlund+2021,
      author = "K. Holmlund and J. Grandell and J. Schmetz and R. Stuhlmann and B. Bojkov and R. Munro and M. Lekouara and D. Coppens and B. Viticchie and T. August and B. Theodore and P. Watts and M. Dobber and G. Fowler and S. Bojinski and A. Schmid and K. Salonen and S. Tjemkes and D. Aminou and P. Blythe",
      title = "Meteosat Third Generation (MTG): Continuation and Innovation of Observations from Geostationary Orbit",
      journal = "Bulletin of the American Meteorological Society",
      year = "2021",
      publisher = "American Meteorological Society",
      address = "Boston MA, USA",
      volume = "102",
      number = "5",
      doi = "10.1175/BAMS-D-19-0304.1",
      pages=      "E990 - E1015",
      url = "https://journals.ametsoc.org/view/journals/bams/102/5/BAMS-D-19-0304.1.xml"
}

@article{Bessho+2016,
  title={An Introduction to Himawari-8/9 - Japan's New-Generation Geostationary Meteorological Satellites},
  author={Kotaro Bessho and Kenji Date and Masahiro Hayashi and Akio Ikeda and Takahito Imai and Hidekazu Inoue and Yukihiro Kumagai and Takuya Miyakawa and Hidehiko Murata and Tomoo Ohno and Arata Okuyama and Ryo Oyama and Yukio Sasaki and Yoshio Shimazu and Kazuki Shimoji and Yasuhiko Sumida and Masuo Suzuki and Hidetaka Taniguchi and Hiroaki Tsuchiyama and Daisaku Uesawa and Hironobu Yokota and Ryo Yoshida},
  journal={Journal of the Meteorological Society of Japan. Ser. II},
  volume={94},
  number={2},
  pages={151-183},
  year={2016},
  doi={10.2151/jmsj.2016-009}
}

@article { TeresakiCorrelatedRMatrixLETKF,
      author = "Koji Terasaki and Takemasa Miyoshi",
      title = "Including the Horizontal Observation Error Correlation in the Ensemble Kalman Filter: Idealized Experiments with NICAM-LETKF",
      journal = "Monthly Weather Review",
      year = "2024",
      publisher = "American Meteorological Society",
      address = "Boston MA, USA",
      volume = "152",
      number = "1",
      doi = "10.1175/MWR-D-23-0053.1",
      pages=      "277 - 293",
      url = "https://journals.ametsoc.org/view/journals/mwre/152/1/MWR-D-23-0053.1.xml"
}

@article{schraff2016,
author = {Schraff, C. and Reich, H. and Rhodin, A. and Schomburg, A. and Stephan, K. and Periáñez, A. and Potthast, R.},
title = {Kilometre-scale ensemble data assimilation for the COSMO model (KENDA)},
journal = {Quarterly Journal of the Royal Meteorological Society},
volume = {142},
number = {696},
pages = {1453-1472},
doi = {https://doi.org/10.1002/qj.2748},
url = {https://rmets.onlinelibrary.wiley.com/doi/abs/10.1002/qj.2748},
eprint = {https://rmets.onlinelibrary.wiley.com/doi/pdf/10.1002/qj.2748},
year = {2016}
}

@article{matricardi2005,
  author={Matricardi, M.},
  year={2005},
  title={The inclusion of aerosols and clouds in RTIASI, the ECMWF fast radiative transfer model for the infrared atmospheric sounding interferometer},
  journal={ECMWF Tech. Memo.},
  volume={474},
  pages={53 pp.}
}

\end{document}